\documentclass[journal] {IEEEtran}
\usepackage{graphicx}
\usepackage{epstopdf}
\usepackage{color}
\usepackage[fleqn]{amsmath}
\usepackage{bm}
\usepackage{subfigure}
\newcommand{\beq}{\begin{equation}}
\newcommand{\eeq}{\end{equation}}
\newcommand{\dis}{\displaystyle}
\newcommand{\beqa}{\begin{eqnarray}}
\newcommand{\eeqa}{\end{eqnarray}}
\newcommand{\bcen}{\begin{center}}
\newcommand{\ecen}{\end{center}}
\newcommand{\blef}{\begin{flushleft}}
\newcommand{\elef}{\end{flushleft}}
\newcommand{\barray}{\begin{array}}
\newcommand{\earray}{\end{array}}

\begin{document}

%
% paper title
\title{Performance Analysis of OFDM with Peak Cancellation Under EVM and ACLR Restrictions}

\author{Tomoya Kageyama~\IEEEmembership{Student member,~IEEE}, 
Osamu Muta,~\IEEEmembership{Member,~IEEE}, \\
and Haris Gacanin,~\IEEEmembership{Senior Member,~IEEE} 
%\thanks{Manuscript received December **, 2015; revised ** ** , ****.}
\thanks{T. Kageyama is with Graduate School of Information Science and Electrical Engineering, 
	Kyushu University in Japan. 
	O. Muta is with Center for Japan-Egypt Cooperation in Science and Technology, 
Kyushu University in Japan (e-mail: muta@ait.kyushu-u.ac.jp, muta@ieee.org). 
H. Gacanin is with Customer Experience Division of Alcatel-Lucent in Belgium.}
%}
\thanks{$^*$This work was presented in part at the 2015 IEEE International Symposium 
on Personal Indoor and Mobile Radio Communications (IEEE PIMRC 2015)~\cite{22}. }
}

%\markboth{IEEE TRANSACTIONS ON ,~Vol.~**, No.~**,~*****~2015}{Shell 
%\MakeLowercase{\textit{et al.}}: Performance Analysis of OFDM with Adaptive Peak Cancellation}

\maketitle

\begin{abstract}

This paper presents performance analysis of an adaptive peak cancellation method to reduce the high peak-to-average power ratio (PAPR) for OFDM systems, while keeping the out-of-band (OoB) power leakage as well as an in-band distortion power below the pre-determined level. 
In this work, the increase of adjacent leakage power ratio (ACLR) and error vector magnitude (EVM) are estimated recursively using the detected peak amplitude. 
We present analytical framework for OFDM-based systems with theoretical bit error rate (BER) representations and detection of optimum peak threshold based on predefined EVM and ACLR requirements. 
Moreover, the optimum peak detection threshold is selected based on the oretical design to maintain the predefined distortion level. 
Thus, 
their degradations are automatically restricted below the pre-defined levels which correspond to target OoB radiation. 
We also discuss the practical design of peak-cancellation (PC) signal with target OoB radiation and in-band distortion 
through optimizing the windowing size of the PC signal.
Numerical results 
show the improvements with respect to both achievable bit error rate (BER) 
and PAPR with the PC method 
in eigen-beam space division multiplexing (E-SDM) systems 
under restriction of OoB power radiation.
It can also be seen that the theoretical BER shows good agreements with simulation results.  
\end{abstract}

\IEEEpeerreviewmaketitle

\section{Introduction}
%\PARstart{M}{ultiple}-input multiple-output (MIMO) technology is considered as an effective approach to increase spectrum efficiency in wireless communications~\cite{1}. 
\PARstart{F}{uture} wireless communication systems require robust communication over a frequency-selective fading channel~\cite{2}, such as orthogonal frequency division multiplexing (OFDM) with multi-input multi-output (MIMO) technologies. One of the technical issues in MIMO-OFDM is the reduction of peak-to-average power ratio (PAPR). 
 
%Hence, a highly linear and power-inefficient amplifier has to be considered in order to avoid signal distortion and out-of-band (OoB) radiation~\cite{3}.

Existing PAPR reduction techniques can be categorized into probabilistic-based approach~\cite{4}-\cite{14}, coding-based~\cite{15}-\cite{17}, and limiter (deliberate clipping)~\cite{18}-\cite{Compand}. 
Contemporary communication systems rely on a simple PAPR reduction technique without any additional processing at the receiver end. Deliberate clipping and filtering (C{\&}F)\cite{18}-\cite{CandF-5}\cite{CandF-1}-\cite{CandF-3} is an attractive technique from the viewpoint of its simple implementation, but it introduces nonlinear degradations. In C{\&}F, filtering is used to remove OoB radiation, but it causes the re-growth of signal amplitude after filtering. Other related approaches such as peak windowing, peak cancellation, and companding (e.g.,\cite{21}-\cite{22}\cite{Compand}) have also been investigated. In principle, C{\&}F and its simplified versions produce nonlinear distortion that may be measured by using error vector magnitude (EVM) and adjacent channel leakage power ratio (ACLR). To cope with them, initial studies of peak cancellation under out-of-band radiation has been presented in~\cite{22b},\cite{22}. However, in these works, the peak detection threshold level is empirically determined and also the optimum peak detection threshold and bit error rate (BER) analysis are not theoretically given.   
{From the practical system design point of view, 
they should be kept below a pre-defined optimum threshold. 
An analytical} evaluation of their impacts on transmission system design is {an important study item}. This is even more important since the bit error rate (BER) performance of the pre-coded OFDM system is highly sensitive to nonlinear degradations.

Main contributions are as follows. 
\begin{itemize}
\item 
Firstly, we present a performance analysis of an adaptive peak cancellation method to keep EVM and ACLR below permissible level 
for multi-input multi-output (MIMO)-OFDM system. 
In this method, an amplitude that exceeds a given threshold is suppressed repeatedly 
by efficient design of peak cancellation (PC) signal, while optimizing the system performance for pre-defined ACLR and EVM. 
We present an efficient distortion estimation method for linearly precoded MIMO-OFDM, 
where the increases of ACLR and EVM are estimated recursively 
using the detected peak amplitude, respectively. 
We confirm that their degradations are restricted below the pre-defined levels 
which correspond to OoB radiation and the level of degradation per subcarrier. 
\item
Secondly, differently from a primitive version of our proposed method in~\cite{22}$^*$, 
this paper presents an analytical framework for OFDM-based systems 
with theoretical BER representations and detection of 
optimum peak threshold  (i.e., theoretically achieved PAPR lower bound) 
based on EVM and ACLR. 
In this framework, 
the optimum peak detection threshold is selected based on 
a theoretical design to achieve the predefined distortion 
level. In other words, achieved PAPR lower bound at pre-defined ACLR and EVM is  theoretically given as the optimum threshold level. 
\item 
In addition, 
we present theoretical BER representations of OFDM with the peak cancellation under EVM and ACLR restrictions 
for single antenna and multi-antenna systems, respectively. 
We clarify that signal distortion due to the peak cancellation is approximated with a  random variable which 
follows Gaussian distribution whose variance and average power is determined based on the optimum peak detection threshold. 
We confirm that achievable PAPR is minimized using the proposed framework 
comparable to those in repeated C\&F method, while restricting 
the distortions within pre-defined levels. 
\item 
We also discuss the practical design of peak-cancellation signal, 
where 
\textcolor{black}{achievable OoB radiation and in-band distortion can be adjusted 
by optimizing the windowing size of the PC signal.} 
%achievable OoB radiation (ACLR) and in-band distortion (EVM) can be adjusted 
%by optimizing the windowing size of the PC signal. 
%
We evaluate and discuss the advantage of our designed method in terms of BER, complementary cumulative distribution function (CCDF) of PAPR as well as computational complexity 
for precoded MIMO-OFDM systems under the restriction of OoB power radiation.
Numerical results clarify the effectiveness of our proposed peak cancellation. 
\end{itemize}
%

%The paper is organized as follows. 
%Section~\ref{section2} describes system model, 
%while the proposed peak-cancellation method is developed in Sect.~\ref{section3}. 
%The results are given in Sect.~\ref{section4}. 
%Section~\ref{section5} concludes the paper.

\section{Related works}
Limiter based PAPR reduction techniques have been investigated for OFDM systems 
in the literature. 
In particular, 
various C\&F based approaches 
are presented such as in~\cite{CandF-5}\cite{CandF-1}-\cite{Compand}. 
In~\cite{CandF-5}, 
an adaptive selection method of peak detection threshold is proposed to achieve fast convergence in repeated C\&F. 
This method is effective in reducing the required number of iterations in C\&F. 
However, in-band distortion due to clipping is not restricted below a pre-defined level.  
In~\cite{CandF-1}, 
an optimized filtering method is proposed for repeated C\&F in which filter characteristic is optimized to minimize  in-band distortion (i.e., EVM) under PAPR constraint while limiting the OoB radiation. Using this method, the required number of repetitions is reduced compared with traditional repeated C\&F. 
In~\cite{CandF-4}, 
a modified repeated C\&F method is presented where the 
clipped signal is optimized by minimizing the increase of in-band distortion at each clipping iteration under PAPR constraint. This method is also effective in reducing the required number of iterations in repeated C\&F. 
However, the above methods still require repeated filtering operations to limit OoB radiation which results in increased computational complexity. 
In~\cite{CandF-2}, 
a simplified C\&F technique using a neural network is proposed. 
In this method, band-limited clipped signal is approximately generated with a learning-based approach without actual filtering process. 
Thus, the required complexity to reduce PAPR is reduced. However, the bit error rate is degraded due to inaccurate approximation as modulation order increases. 
Thus, more complicated learning method may be needed for signals with higher-order modulation. 
In~\cite{CandF-3}, 
a repeated C\&F method that adaptively determines the clipping threshold is proposed. 
Although this method does not require a pre-defined threshold, in-band distortion is not restricted below a pre-defined level. 
In~\cite{Compand}, 
a new limiter based companding function was proposed  to 
suppress peak amplitude effectively. 
However, companding transformation causes OoB radiation
due to nonlinearity of companding function. 

Unlike the above approaches, in this paper, we designed an effective peak cancellation method that automatically restricts EVM and ACLR below the optimum predefined level, while reducing the peak amplitude below the threshold level with lower complexity than the conventional repeated C\&F.

\section{Mathematical Signal Representation~\label{section2}}

\begin{figure}[t]
%\bcen
\includegraphics[scale=0.35]{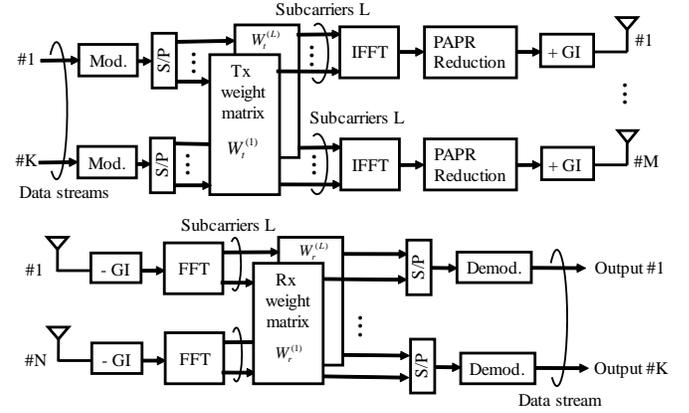}
%\ecen
%\hspace{5mm}
\caption{System block diagram.}
\label{fig:ofdm}
\end{figure}

\subsection{System Model}
Figure~\ref{fig:ofdm} illustrates an E-SDM OFDM system model, 
where $M$, $N$, and $K$ 
denote the number of transmit antennas, receive antennas, and data streams, respectively.
$\bm{W}_t^l$ and $\bm{W}_r^l$ denote precoding and post-coding matrices on the $l$-th subcarrier, respectively, 
where {$l=1,\cdots, L$, and $L$} denotes the number of subcarriers.
Here, the transmit data vector of the $l$-th subcarrier $\bm{x}^l=[x_1^l, ...,x_k^l,...,x_K^l]^T$ is multiplexed 
by the precoding matrix $\bm{W}_t^l = [\bm{w}_{t1}^l, ...,\bm{w}_{tk}^l, ...\bm{w}_{tK}^l]$,
where $\bm{w}_{tk}^l = [w_{tk1}^l, ...,w_{tkm}^l, \dots,w_{tkM}^l]^T$ is the $l$-th column vector of $\bm{W_t^l}$ 
for the $k$-th data stream and superscript 
$T$ stands for transposed matrix. 
\textcolor{black}{
	In E-SDM, left singular vector and right singular vector of the channel matrix are used as precoding and post-coding matrices, respectively. }
{$\bm{H}^l$ stands for $N \times M$ matrix of the $l$-th subcarrier defined as}
\begin{eqnarray}
\bm{H}^l = \left(
\begin{array}{cccc}
h_{11}^l & \ldots & h_{1M}^l \\
\vdots & h_{nm}^l & \vdots \\
h_{N1}^l & \ldots & h_{NM}^l
\end{array}
\right),
\end{eqnarray}
where $h_{nm}$ denotes impulse response of the path, 
Here, the $m$ and $n$ denote the transmit antenna index and the receive antenna index, respectively. 
\textcolor{black}{Using  with singular value decomposition (SVD), ${\bm H}^l$ can be decomposed into 
%\textcolor{black}{${\bm H}^l$ can be written as follows using singular value decomposition. 
\begin{eqnarray}
{\bm H}^l = {\bm U}_l \Sigma_l {\bm V}_l^H, 
\end{eqnarray}
where ${\bm U}_l$ and ${\bm V}_l^H$ are left and right singular vector of ${\bm H}^l$, and $(\cdot)^H$ means complex conjugate operation. $\Sigma_l = {\rm diag}(\sqrt{\lambda_1^l}, \sqrt{\lambda_2^l}, ..., \sqrt{\lambda_M^l})$ is a diagonal matrix and $\sqrt{\lambda_m^l}$ is the singular value of $m$-th stream. 
Transmit and receive spatial filter are defined as ${\bm W}_t^l = {\bm V}_l$ and 
${\bm W}_r^l = {\bm U}_l^H$. 
}
The precoded QAM data stream is modulated with inverse fast Fourier transform (IFFT) 
and then PAPR reduction technique is applied at each transmits antenna. 
The guard interval (GI) is added to every symbol to remove inter-symbol interference.
Perfect channel estimation is assumed.

After removing GI and carrying out FFT processing, 
the received signal is multiplied by the post coding matrix 
$\bm{W_r}^l = [\bm{w_{r1}}^l, ...,\bm{w_{rk}}^l, ...\bm{w_{rK}}^l]^T$,
where  $\bm{w}_{rn}^l =[w_{rn1}^l, ...,w_{rnk}^l, ...,w_{rnK}^l]^T$ 
denotes the $n$-th post-coding vector of the $l$-th subcarrier. 
Hence,
the de-multiplexed signal vector $\bm{y}^l = [y_1^l, ...y_k^l, ...y_K^l]^T$ of the $l$-th subcarrier is given as
\begin{eqnarray}
{\bm y}^l &=& {\bm W}_r^l({\bm H}^l{\bm W}_t^l{\bm x}^l + {\bm n}^l) \nonumber \\
&=& {\bm U}_l^H({\bm H}^l{\bm V}_l{\bm x}^l + {\bm n}^l) \nonumber \\
&=& {\bm U}_l^H({\bm U}_l\Sigma_l{\bm V}_l^H{\bm V}_l{\bm x}^l + {\bm n}^l) \nonumber \\
\label{eq:esdm2}
%&=& {\rm diag}(\sqrt{\lambda_1^l}, \sqrt{\lambda_2^l}, ..., \sqrt{\lambda_M^l}){\bm x}^l + {\bm U}_l^H {\bm n}^l, 
&=& \Sigma_l{\bm x}^l + {\bm U}_l^H {\bm n}^l, 
\end{eqnarray}
%where ${\bm U}_l$ and ${\bm V}_l$ are unitary matrix and 
%\begin{equation}
%\bm{y}^l = \bm{W}_r^l\bm{H}^l\bm{W}_t^l\bm{x}^l + \bm{W}_r^l\bm{n}
%= {\rm{diag}} (\bm{\lambda}^l)\bm{x}^l + \bm{W}_r^l\bm{n},
%\label{eq:esdm2}
%\end{equation}
where ${\bm U}_l$ and ${\bm V}_l$ are unitary matrices and 
$\bm{n}^l=[n_1, ...n_n, ...,n_N]^T$ is an additive white Gaussian \textcolor{black}{noise} (AWGN) vector 
at each receive antenna. %${\rm diag}({\bm \lambda}^l) = {\rm diag}[\lambda_1^l, ...\lambda_k^l, ...\lambda_K^l]$ is 
%diagonal matrix with elements being an eigen value of channel matrix $\bm{H}^l$.

\subsection{Definitions of ACLR and EVM}
In this paper, we evaluate the amount of OoB radiations as ACLR which is defined as 
\begin{eqnarray}
\mbox{ACLR}=\int^{}_{ {\rm f_U, \ f_L} }\frac{S(\omega)}{S_t}d\omega,
\label{ACLR}
\end{eqnarray}
Here, {$S_t = \frac{1}{2} E\left[|x^l_k|^2\right] $ denotes the average power of the transmit signal 
and $S(\omega)$ is the power spectral density of the transmitted signal. 
%If we assume that the frequency bandwidth of complex OFDM signal is set to $(-\frac{L}{2}+1)f_0 \sim \frac{L}{2}f_0$, 
%OoB radiation spectral power is measured at $f_U=(\frac{L}{2}+3)f_0 \sim (\frac{3L}{2}+2)f_0$ and $f_L=(-\frac{3L}{2}-1)f_0 \sim (-\frac{L}{2}-2)f_0$, 
%where $L$ is the number of subcarriers. 
%
{%$f_0=\frac{1}{T}$, 
%where $T$ denotes OFDM symbol duration.} 
Let $f_U$- and $f_L$ denote the measured ACLR at an upper and lower band, respectively.
%We denote ACLR measured at upper band and lower band as $f_U$- and $f_L$-band, respectively.
The permissible maximum ACLR is set to $-50$~dB for transmitting signals at every antennas. 

We also evaluate in-band distortion by measuring EVM which is defined as: 
\begin{equation}
\mbox{EVM}=\sum^{L/2}_{l=-L/2+1}{|X_l-\tilde{X}_l|^2}/\sum^{L/2}_{l=-L/2+1}{S_t[l]},
\label{evm}
\end{equation}
where $S_t[l]$ is the average power of the $l$-th subcarrier signal. 
$X_l$ and $\tilde{X}_l$ 
denote {the complex signals at the $l$-th subcarrier (after FFT operation)} 
without and with PAPR reduction, 
respectively. 
%Thus, the amount of in-band distortion due to PAPR reduction can be evaluated by measuring EVM. 
In E-SDM case, 
$X_l$ and $\hat{X}_l$ are defined as complex signals and 
{replicated at} the $l$-th subcarrier of each eigen-channel by Eq.~(\ref{eq:esdm2}).

\section{Adaptive Peak-cancellation Technique\label{section3}}
{Figure~\ref{fig:pcblock} shows the block diagram of our designed peak cancellation 
which consists of five steps, i.e., 
PC signal generation, selection of the peak detection threshold, peak detector, 
distortion estimation, and PC signal scaling, 
where $x^{(m, i)}(t)$ denotes input signal at $m$-th antenna at $i$-th repetition. 
The first two blocks (``PC signal generation'' and ``selection of the peak detection threshold'') 
are carried out beforehand. 
In each transmission frame, 
whenever the maximum amplitude $x_{max}^{(m,i)}$ exceeding the threshold $A_{th}$ 
is detected at the peak detector, 
the distortion estimation is carried out using a recursive method, 
and it decides whether both ACLR and EVM are below 
their given values or not (i.e., binary decision; ``Y'' or ``N''). 
An average EVM value and the maximum ACLR value over all transmit antennas are estimated, 
and then if the estimated values are below {the pre-defined} values, 
the amplitude of the PC signal is scaled to $A^{(m,i)}_p=x_{max}^{(m,i)} - A_{th} $. 
Then, the scaled PC signal is added to reduce $x_{max}^{(m,i)}$ to $A_{th}$, 
Otherwise, the peak cancellation procedure stops. 
The details of each {step} are explained below.}

\subsection*{(Step 1) Peak Cancellation Signal Generation\label{section3a}}
Whenever the maximum signal amplitude exceeds a given threshold level $A_{th}$,
the maximum peak is suppressed by adding a PC signal. 
Here, the PC signal is a scaled OFDM symbol whose subcarriers are added 
up to be in-phase at a given symbol time instant. 
The PC signal is generated by scaling the following basic function $g(t)$ as 
\beq
g(t)=\frac{1}{L}\sum_{l=-L/2+1}^{L/2} p_l (t) e^{j\omega_l t}, 
\label{pcsignal}
\eeq
where {$p_l(t)$ is the transmit pulse at the $l$-th subcarrier. 
In this paper, we assume that $p_l(t)$ is the same rectangular pulse on all subcarriers 
whose amplitude is unity. }
The time-domain waveform of $g(t)$ is truncated 
by a windowing function $w(t)$; the truncated version of $g(t)$ is given as 
\beq
g'(t)=w(t)g(t).
\label{g'(t)}
\eeq
We use the following windowing function to truncate the peak cancellation signal waveform: 
%\vspace{-2mm}
\begin{eqnarray}
%\hspace{-2mm}
%\scriptstyle{
%\begin{displaymath}{
w(t)=\left\{
\begin{array}{ll}
0 & (T_2 < |t|) \\
\displaystyle \frac{1}{2}+\frac{1}{2}\cos{\frac{\pi(|t|-T_1)}{T_2-T_1}} & (T_1 < |t| \le T_2) \\
1 & (|t| \le T_1),
\end{array}
\right.
%}
%\nonumber
\label{window}
\end{eqnarray}
where $0 < T_1 \le T_2$.
Here, $T_1$ is a design parameter to optimize distortions appeared after peak cancellation. 
{$T_2$ denotes window size of $w(t)$. }
%Note that Hanning window function corresponds to case using $T_1=0$ in Eq.~(\ref{window}). 
% 
The truncated PC signal waveform and its frequency spectrum of $g'(t)$ are illustrated in
Figs.~\ref{fig:PCsignal}~(a) and (b), respectively. 
Here, parameters of windowing function are given as $(T_1, T_2)=(T/8, T/4)$.
The PC signal explicitly exhibits high peak amplitude which is utilized 
to reduce the PAPR of OFDM.

\begin{figure}[t]
%	\bcen
	\hspace{5mm}\includegraphics[scale=0.30]{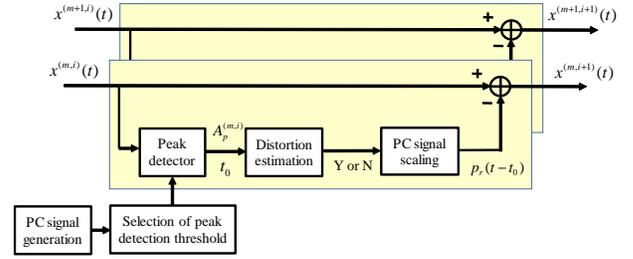}
%	\ecen
\vspace{5mm}
	\caption{Block diagram of the proposed method, 
			where $x^{(m)}(t, i)$ denotes the transmit signal at $m$-th antenna 
			after adding $i$-th PC signal.}
	\label{fig:pcblock}
\end{figure}
\begin{figure}[t]
	\begin{center}
		\subfigure[Time-domain waveform.]{
			\includegraphics[scale=0.65]{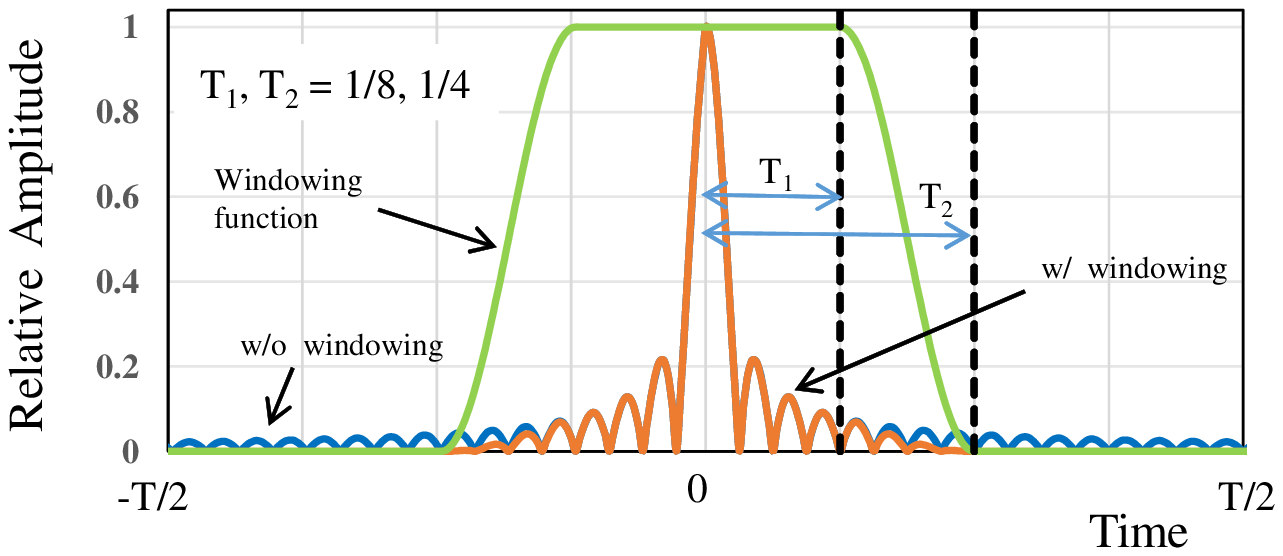}} %scale=0.29
		\label{fig:PCsignalt}
		\subfigure[Frequency-domain spectrum.]{
			\includegraphics[scale=0.65]{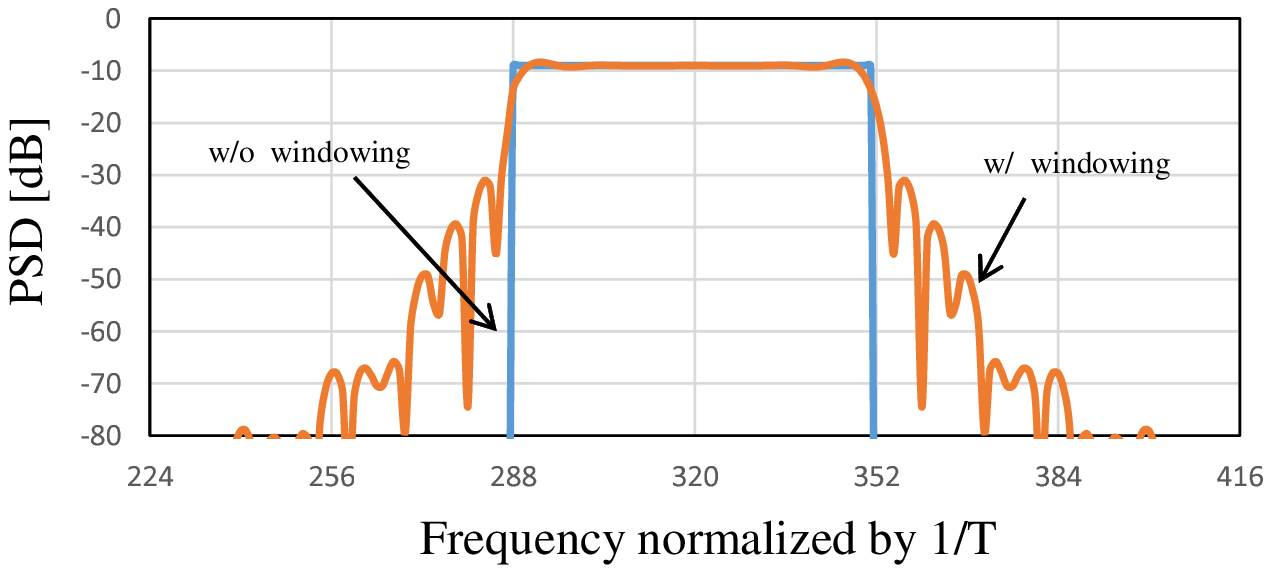}} %scale=0.29
		\label{fig:PCsignalf}
		\caption{{An example of peak cancellation (PC) signal.}}
	\label{fig:PCsignal}
	\end{center}
\end{figure}
\subsection*{(Step 2) Selection of Peak Detection Threshold\label{section3c}}
In the proposed method, 
the optimum peak detection threshold should be selected so that the signal amplitude is 
suppressed below the threshold level, 
while EVM and ACLR are kept below the pre-defined values. 
This subsection describes the optimum threshold selection.
Figure~\ref{fig:waveform} illustrates an example of an OFDM signal waveform. 
In this figure, 
red and blue lines show the signal with and without the proposed peak cancellation, respectively. 
The green-shaded area corresponds to signal amplitude exceeding the peak detection 
threshold $A_{th}$.
As illustrated in Fig.~\ref{fig:waveform}, 
adding the PC signal distorts the OFDM signal and increases both EVM and ACLR. 
\begin{figure}[t]
%\bcen
\hspace{5mm}\includegraphics[scale=0.30]{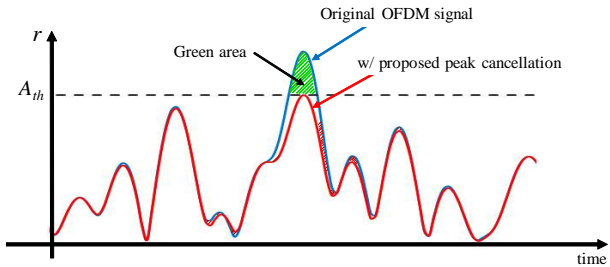}
%\ecen
%\vspace{5mm}
\caption{Illustration of the peak cancellation effect.}
\label{fig:waveform}
\end{figure}
\begin{figure}[t]
%\bcen
\hspace{15mm}\includegraphics[scale = 0.25]{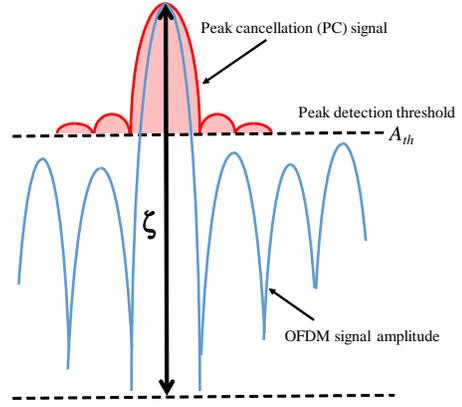}
%\ecen
%\vspace{5mm}
\caption{Modeling of signal distortion due to PC signal addition.}
\label{fig:model}
\end{figure}

Figure~\ref{fig:model} illustrates the relation between the scaling factor of the PC signal 
and the peak detection threshold, 
where $\zeta$ and $A_{th}$ denote the maximum amplitude of the OFDM signal and the peak detection threshold, respectively. 
%
%Probability density function of instantaneous power of OFDM signal $\zeta^2$ is denoted as 
%$r(\zeta^2)$. 
%
As illustrated in Fig.~\ref{fig:model}, 
when the maximum amplitude $\zeta$ exceeds the threshold level $A_{th}$, 
OFDM signal amplitude is reduced to the threshold level by 
adding the PC signal whose amplitude is scaled to $\zeta-A_{th}$.
Since the scaled PC signal is given as $(\zeta-A_{th})g'(t)$, 
the energy of the scaled PC signal is calculated as 
\begin{eqnarray}
E(\zeta - A_{th}) \!\!\!\!\! &=& \!\!\!\!\!
\int_{-T_2}^{T_2} (\zeta -A_{th})^2g'^2(t) dt \nonumber \\
\!\!&=& \!\!\!\!\!(\zeta -A_{th})^2 { \int_{-\infty}^{\infty} G'^2(\omega) d \omega } \nonumber \\
\!\!&=& \!\!\!\!\!(\zeta -A_{th})^2 \left( {\int_{f_{{\rm in}}} G'^2(\omega) d \omega} 
+ {\int_{f_{{\rm out}}} G'^2(\omega) d \omega} \right) \nonumber \\
\!\!&\equiv& \!\!\!\!\!\! (\zeta -A_{th})^2(E_i + E_o), 
\end{eqnarray}
where 
%$f_{in}$ and $f_{out}$ denote bandwidths to measure inband distortion (EVM) and 
%OoB radiation (ACLR), respectively.
\textcolor{black}{$f_{in}$ and $f_{out}$ denote bandwidths to measure EVM and 
ACLR, respectively.} 
Here, $G'(\omega) = \int_{-\infty}^{\infty} g'(t) e^{-j\omega t} dt$. 
%and $T_2$ is a half of window size. 
%
$E_i$ and $E_o$ denote in-band power and out-of-band power of PC signal, respectively. 
Statistical distribution of instantaneous power of the OFDM signal 
follows exponential distribution as 
\begin{eqnarray}
r(\zeta^2) = \frac{1}{\sigma^2} {\rm exp} \left( -\frac{\zeta^2}{\sigma^2} \right), 
\end{eqnarray}
where $\zeta^2$ and $S_t=\sigma^2$ denotes instantaneous power 
and average power of OFDM signal, respectively.
Thus, the average signal distortion due to peak cancellation 
is calculated as 
\begin{eqnarray}
S_d  \! &=& \int_{A_{th}}^{\infty} E(\zeta-A_{th}) r(\zeta^2) d\zeta^2 \nonumber\\
&=& (E_i + E_o) \int_{A_{th}}^{\infty} (\zeta-A_{th})^2 r(\zeta^2) d\zeta^2. %\nonumber \\
%&\approx& E_i \int_{A_{th}}^{\infty} (\zeta - A_{th})^2 r(\zeta^2) d\zeta^2.
\end{eqnarray}
The in-band distortion power is given as 
\begin{equation}
S_{in}= E_i  \int_{A_{th}}^{\infty} (\zeta-A_{th})^2 r(\zeta^2) d\zeta^2.
\end{equation}
%%
%Let $S_i$ and $S_t$ denote inband distortion and average power, respectively. 
Assuming $E_i \gg E_o$, 
%%\begin{eqnarray}
the normalized signal distortion is restricted as  
\beq
%10&{\rm log}_{10}&\left( 
\frac{S_{d}}{S_{t}} 
\approx 
\frac{S_{in}}{\sigma^2}
%=
%\frac{E_i}{\sigma^2} \int_{A_{th}}^{\infty}(\zeta - A_{th})^2r(\zeta^2) d\zeta 
\le \frac{e_r}{\sigma^2}, 
\eeq
%%\end{eqnarray}
where the average power of the signal is $S_{t}=\sigma^2$ 
and $\frac{e_r}{\sigma^2}$ denotes the pre-determined value of the maximum acceptable EVM. 
Hence, the minimum (optimum) peak detection threshold $A_{th}^o$ that meets the EVM requirement is given as 
\beq
%10{\rm log}_{10} \left[
\int_{A^o_{th}}^{\infty}(\zeta - A^o_{th})^2r(\zeta^2) d\zeta 
%\right]. 
= \frac{e_r}{E_i} . 
\label{eqevm_est}
\eeq
Here, $(A^o_{th})^2$ corresponds to the peak power of the OFDM signal after peak cancellation. 
Using Eq.(\ref{eqevm_est}), the optimum threshold $A^o_{th}$ to achieve 
pre-determined EVM value can be theoretically obtained. 
%
%%%
%The relation between the optimal threshold value $A_{th}^o$ and the required EVM in dB $e_r$ is as follows.
%\begin{eqnarray}
% 10{\rm log}_{10} \left[\frac{E_i}{\sigma^2} \int_{A^o_{th}}^{\infty}(\zeta - A^o_{th})^2r(\zeta^2) d\zeta \right] = e_r {\rm \ in \ dB}. 
%\label{eqevm_est}
%\end{eqnarray}
%The optimal threshold level $A_{th}^o$ can be determined so as to meet Eq.(\ref{eqevm_est}). 

\subsection*{(Step 3) Peak Detection}
{
In Fig.~\ref{fig:pcblock}, 
the remaining three steps (i.e., ``peak detection'', ``distortion estimation'', 
and ``PC signal scaling'') are repeated until the maximum value is below the threshold value 
unless ACLR or EVM exceeds their pre-defined values. 
At the first step, 
when the peak amplitude of the OFDM signal at $m$-th antenna 
exceeds the selected peak detection threshold at time instance $t=t_0$, 
{the difference between the detected peak amplitude and the detection threshold, 
$A_p^{(m, i)}$, is calculated.} 
In the next step, 
the increase of ACLR and EVM is estimated using the detected value $A_p^{(m,i)}$. 
}

\subsection*{{(Step 4) Distortion Estimation\label{section3b}}}
In multi-stream transmission in MIMO-OFDM, 
the peak cancellation is carried out to suppress the peak amplitude below the peak detection threshold under constraints of EVM and ACLR, 
where EVM and ACLR requirements are defined as an average value and the maximum value over all antennas, respectively. 

%to restrict 
%an average EVM value and the maximum ACLR value over all antennas 
%below a permissible level, while suppressing the peak amplitude below the peak detection threshold. 
% 
%
Since the truncated signal $g'[s]=g'(s\Delta t)$ has out-of-band spectrum,  
it is clear that out-of-band radiation and in-band distortion appears by adding $g'[s]$to the transmit signal, 
where $\Delta t$ denotes sampling interval. 
Here, 
$$G'[l] \mathcal{F}[g'[s]]=W[l]{\ast}G[l],$$
where $\mathcal{F}$ and $\ast$ denote discrete Fourier transform 
and convolution operator, respectively. 
$G[l]$ and $W[l]$ are frequency spectrum of the PC signal $g[s]$ and 
the window function $w[s]$, respectively.

{Let $\Delta p_{o}$ and $\Delta p_{in}$ denote OoB signal power and 
in-band signal power of $g'(t)$. 
Note that 
$\Delta p_{o}$ and $\Delta p_{in}$ are known values (calculated beforehand).} 
Hence, when $G'[l]$ is added to the signal to cancel the peak,  
the in-band distortion and OoB radiation are increased by 
\beq
\left\{
\begin{array}{ll}
\Delta p_{in} =& \dis\sum_{l=-L/2+1}^{L/2} |G'[l]|^2, \\
\Delta p_{o} 
= & 
% \left(
{\dis{\sum_{l=L/2+2}^{3L/2+1}|G'[l]|^2}+{\dis\sum_{l=-3L/2}^{-(L+2)/2} |G'[l]|^2}}. 
%\right). 
\end{array}
\right.
\label{Eqp_o}
\eeq

Let the total transmission power $S_t$ be constant, i.e., {$S_t=\sum_{m=1}^M S^{(m)}$, 
where $S^{(m)}$} is the average signal power at the $m$-th transmit antenna. 
Let $|A_p^{(m, i)}|$ denote the difference between the threshold value 
and the $i$-th peak amplitude $x^{(i)}_{max}$ at the $m$-th antenna.
EVM increase is expressed 
{as 
	$
	\displaystyle\frac{1}{S_t} |A_p^{(m, i)}|^2\Delta p_{in} 
	$
}
when the PC signal is added to suppress the $i$-th peak amplitude $x^{(i)}_{max}$. 
%Thus, the $i$-th peak amplitude $x^{(i)}_{max}$ can be suppressed 
%by adding the PC signal which increases the EVM 
%{by 
%$
%\displaystyle\frac{1}{S_t} |A_p^{(m, i)}|^2\Delta p_{in}.
%$
%}

We calculate an average EVM value overall antennas as
\beq
\Delta\varepsilon_e^{(i)}
=
\sum_{m=1}^M \frac{1}{S_t} |A_p^{(m, i)}|^2\Delta p_{in}
=\frac{\Delta p_{in}}{S_t} \sum_{m=1}^M|A_p^{(m, i)}|^2, 
\label{eq10}
\eeq
where {$\Delta p_{in}$ denotes the pre-determined constant}.
The averaged EVM value $\varepsilon_e^{(i)}$ is recursively calculated as
\begin{equation}
\varepsilon_e^{(i)}=\varepsilon_e^{(i-1)}+ \Delta \varepsilon_e^{(i-1)}. 
\label{est_eq2}
\end{equation}

In order to restrict OoB radiation below the permissible value, 
we propose to estimate the instantaneous ACLR at each antenna as follows: 
When the $i$-th PC signal is added, the ACLR at each antenna is increased as 
\beq
\Delta\varepsilon_a^{(m, i)}
=
\frac{1}{S_t / M } |A_p^{(m, i)}|^2 \Delta p_{o}
=
\frac{M\Delta p_{o}}{S_t } |A_p^{(m, i)}|^2. 
\label{eq11}
\eeq
Using the above relation, the ACLR $\varepsilon_a^{(m, i)}$ after adding the $i$-th PC signal 
is recursively calculated as 
%.
\begin{equation}
\varepsilon_a^{(m, i)}=\varepsilon_a^{(m, i-1)}+{\Delta \varepsilon_a^{(m, i)}}. 
\label{est_eq}
\end{equation}

Finally, 
the maximum value is calculated as
\beq
{\varepsilon_a^{(i)}}_{max}=f(\varepsilon_a^{(m, i)})_{max:m}, 
\label{eq.evm_m}
\eeq
where $f(\cdot)_{max:m}$ is a function that selects 
the maximum value from possible ones with respect to antenna index $m$. From the practical design point 
of view Eq.~(\ref{eq10}) maybe used to directly calculate the level of degradation per subcarrier by simply reading out the signal amplitude level $A_p^{(m,i)}$ 
{at the transmitter end.}

{
The PC signal scaling and peak cancellation procedure in next step is done 
if both the average EVM $\varepsilon_e^{(i)}$ in Eq.~(\ref{est_eq2}) and 
the maximum ACLR ${\varepsilon_a^{(i)}}_{max}$ in Eq.~(\ref{eq.evm_m})
are less than the pre-defined levels. 
}

\subsection*{(Step 5) PC Signal Scaling and Peak Cancellation}
To carry out the peak cancellation (i.e., the PC signal addition), 
both the averaged EVM values $\varepsilon_e^{(i)}$ and the estimated maximum ACLR ${\varepsilon_a^{(i)}}_{max}$ must be less than permissible values. 
Our proposed scheme can satisfy this condition automatically as explained below. 
First, when the maximum amplitude of the signal $x(t_0)$ exceeds the threshold value,  
the scaled PC signal is expressed as 
\beq
\barray{l}
p_r(t-t_0) \\
= -A_p \dis\frac{1}{L}w(t-t_0)\dis\sum_{l=-L/2+1}^{L/2} p(t-t_0) e^{ j(\omega_l(t-t_0)+\theta_0) } \\
\equiv -A_p e^{j\theta_0} g'(t-t_0),
\earray
\label{Eq1}
\eeq
where
$A_p =|x(t_0)|-A_{th}$.
$\theta_0$ is the phase of the signal $x(t_0)$.
Assume that the maximum amplitude exceeds the peak detection threshold at time instance $t^{(i)}_0$. 
Then, 
the amplitude of the OFDM signal after canceling the $i$-th peak amplitude
is represented as 
\begin{equation}
x^{(i)}(t) = x^{(i-1)}(t)+p_r^{(i)}(t-t^{(i)}_0). 
\end{equation}
The above procedures are continued repeatedly until all amplitudes are suppressed below the peak detection threshold or the number of PC signal additions reaches a maximum number. 
However, if either the estimated ACLR or EVM exceeds the permissible value, the peak cancellation procedure stops before it reaches the maximum number. This stopping criterion guarantees that 
ACLR and EVM never exceed their permissible values. 
\\

\section{BER analysis of OFDM with the designed peak cancellation}
In this section, 
we analyze theoretical BERs of single antenna OFDM and linear precoded (E-SDM) MIMO-OFDM 
with our designed peak cancellation, respectively.

\subsection{Single antenna OFDM}
In this section, first, we consider QPSK-OFDM signal whose subcarrier's I and Q-phases take 
either $A$ or $-A$. 
Let $x_e$ denote a random variable which expresses signal distortion due to peak cancellation. 
Thus, the signal after peak cancellation is given as $\pm A + x_e$. 
%When channel gain is $\beta$, 
We assume that 
%{$x_e$} is a random variable whose PDF, i.e., $p_e(x_e)$, 
$x_e$ follows 
%independent identical 
Gaussian distribution with variance $\sigma_e^2$ and average value $\bar{x}_e$; 
\beq
p_e(x_e)=\frac{1}{\sqrt{2\pi}\sigma_e}\exp{\left( -\frac{(x_e-\bar{x}_e)^2}{2\sigma_e^2}\right)}, 
\eeq
where 
$\sigma_e^2$ and $\bar{x}_e$ are variance and average amplitude of $x_e$. 
Since the average power of the signal is decreased after peak cancellation, 
it can be intuitively seen that $\bar{x}_e$ decreases as $A_{th}$ decreases, 
while $\sigma_e$ increase as $A_{th}$ decreases. 
Hence, the average value of $x_e$ can be expressed with $S_{in}$ as 
\beq
\bar{x}_e \approx %\mu A =
- \alpha \sqrt{S_{in}/L}, 
\eeq
where 
%$\mu$ is a scaling parameter whose range is $0 < \mu \le 1$. 
$L$ denotes the number of subcarriers. 
$\alpha$ is a constant depending on PC signal waveform $g'(t)$ 
given as 
\beq
%a=\frac{\displaystyle\int^\infty_{-\infty} f_c(g'(t))dt} { \displaystyle\int^\infty_{-\infty} |g'(t)| dt}
\alpha =\frac{\displaystyle\int^\infty_{-\infty} f_c({\rm R_e}[g'(t)])dt}
{\displaystyle\int^\infty_{-\infty} |{\rm R_e}[g'(t)] |dt}. 
\label{a}
\eeq
where $f_c(x)$ denotes a clipping function defined as 
\beq
f_c(x)=
\left\{
\begin{array}{cc}
x & x > 0 \\ 
0 & \mbox{otherwise}. 
\end{array}
\right.
%$
\eeq
%\vspace{2mm}
The numerator and the denominator in Eq.(\ref{a}) show integral of positive side PC signal amplitude 
and that of its absolute value, respectively. 
The variance of $x_e$ is given as 
$$
\sigma_e^2
=\langle (A+x_e - \mu A)^2 \rangle 
%=\mu^2 \langle ((A+x_e)/\mu - A)^2 \rangle 
=\mu^2 (S_{in}/L), 
%%= \mu^2 \frac{S_{in}}{L}
%=\langle(\mu x_e -\mu A )^2\rangle 
%= \mu^2 \langle (x_e - A ) ^2 \rangle 
%= \mu^2 \frac{S_{in}}{L}, 
$$
where $\mu=1-\bar{x}_e/A$. 
Here, $\langle x \rangle$ denotes expected value of the variable $x$. 

Let $x_n$ denote additive white Gaussian noise (AWGN) with average level $A$ and 
variance $\sigma_n^2$, i.e., PDF of $x_n$ is given as 
\beq
p_n(x_n, \sigma_n)=
\frac{1}{\sqrt{2\pi }\sigma_n}
\exp{\left(-\frac{x_n^2}{2\sigma_n^2}\right)}.
\eeq
$x_e$ and $x_n$ are Independent and identically distributed (i.i.d.) random variables and therefore 
PDF of mixed variable $x=x_e+x_n$ is given as convolution of $p_x(x)$ and $p_n(x)$: 
%
%\vspace{-2mm}
\begin{eqnarray}
&&
%\hspace{-10mm}
p(x, \sigma_n, \sigma_e, \beta)
%\hspace{-3.5mm} &=&\hspace{-3.5mm} 
=\int_{-\infty}^{\infty} p_e(x-y, \sigma_e, \beta)p_n(y, \sigma_n)dy \nonumber \\
&=& 
%\hspace{-3.5mm} 
\frac{1}{\sqrt{2\pi((\sigma_e^2/\beta)+\sigma_n^2)}}
\exp \left({-\frac{(x-(\bar{x}_e/\sqrt{\beta}))^2}{2((\sigma_e^2/\beta)+\sigma_n^2)}} \right), 
\nonumber
\end{eqnarray}
where $\beta$ denotes channel gain. 
BER for QPSK-OFDM signal after peak cancellation in AWGN condition with channel gain is expressed as 
%\begin{fleqn}[20pt]
%\begin{align*}
\begin{eqnarray}
&\!\!\!\!\!\!\!\!\!\!\! P_b 
\left(A, \sigma_n, \sigma_e, \beta \right) 
= \displaystyle\int^0_{-\infty} p(x-A, \sigma_n, \sigma_e, \beta) dx \nonumber \\
&\!\!\!\!\!\!\!\!\!\!\!= \hspace{-1.0mm} 
\displaystyle\int_0^\infty \hspace{-3.5mm} 
\frac{1}{\sqrt{2\pi((\sigma_e^2/\beta)+\sigma_n^2)}}
\exp \hspace{-1.0mm}
\left({-\frac{(x-(\bar{x}_e/\sqrt{\beta}))^2}{2((\sigma_e^2/\beta)+\sigma_n^2)}} \right)
\hspace{-0.5mm} dx, 
\label{eq_ber}
\end{eqnarray}
%\end{align*} 
%\end{fleqn}
where $A$ denotes I and Q phase signal amplitudes. 

%where $\bar{x}_e$ denotes average power and 
%the variance is given as $\sigma_e^2+\sigma_n^2$. 
%Figure~\ref{mixed_noise} shows an example of distributions of $p_e(x)$, $p_n(x)$, and $p(x)$. 

%
%\begin{figure}[tb] 
%\begin{center}
%\includegraphics[scale=0.33]{fig/16QAM.eps}\\[1mm]
%\end{center}
%\vspace{0mm} 
%\caption{16QAM signal distribution (I-phase or Q-phase).}
%\label{16QAM}
%% \vspace{-2mm} 
%\end{figure}
\begin{figure}[t]
	\begin{center}
		\subfigure[w/o peak cancellation (in the presence of AWGN).]{
			\includegraphics[scale=0.45]{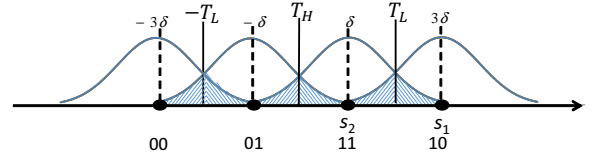} %scale=0.29
		\label{fig:16QAMwo}}
		\subfigure[w/ peak cancellation (in the presence of AWGN and signal distribution).]{
			\includegraphics[scale=0.45]{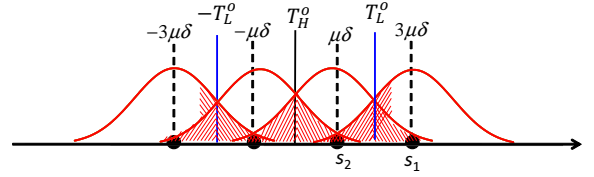} %scale=0.29
		\label{fig:16QAMw}}
		\caption{{16QAM signal distribution (I-phase or Q-phase).}}
	\label{fig:16QAM}
	\end{center}
\end{figure}

The above discussion can be extended to multi-level QAM such as 16QAM and 64QAM. 
The same manner as QPSK case can be used for deriving BER expressions in QAM case 
except that variance of in-band noise is different, e.g., 
for 16QAM and 64QAM cases, their variances $\sigma_e^{16QAM}$ and 
$\sigma_e^{64QAM}$ are respectively given as 
\beq
\sigma_e^{16QAM}=\sigma_e/2, 
\eeq
\beq
\sigma_e^{64QAM}=\sigma_e/4, 
\eeq
where $\sigma_e$ denotes variance in QPSK case. 
More generally, variance for $2^{Q}$ QAM is given as $\sigma_e/Q$
%
%Although the details of derivations for QAM cases are omitted in this paper due to limited pages, 
%their theoretical BER graphs are shown in next section.

Figure~\ref{fig:16QAM} shows I-phase or Q-phase signal distribution of 16QAM at a certain subcarrier 
in the case with and without peak cancellation. 
Error probability of higher order bit $P_{eH}$ is given as 
a probability that signal $S_1$ and $S_2$ exceed threshold $T_H$; 
\beqa
P_{eH} & = &
\frac{1}{2}\frac{1}{\sqrt{2\pi} \sigma_n} 
\left\{
\int_0^\infty \exp\left(-\frac{(x+\delta)^2}{2\sigma_n^2} \right)dx 
\right.
\nonumber \\ 
&+& 
%\frac{1}{2}\frac{1}{\sqrt{2\pi \sigma_n^2}} 
\left.
\int_0^\infty \exp\left(-\frac{(x+3\delta)^2}{2\sigma_n^2} \right)dx
\right\}. 
%\nonnumber
\eeqa
Error probability of lower order bit of 16QAM is given as 
a probability that signal $S_1$ and $S_2$ exceed threshold $T_L$; 
\beqa
P_{eL} & = &
\frac{1}{2}\frac{1}{\sqrt{2\pi} \sigma_n} 
\left\{
\int_{-2\delta}^{2\delta} \exp\left(-\frac{(x+3\delta)^2}{2\sigma_n^2} \right)dx 
\right.
\nonumber\\ 
&+& 
%\frac{1}{2}\frac{1}{\sqrt{2\pi \sigma_n^2}} 
\int_{-\infty}^{-2\delta} \exp\left(-\frac{(x+\delta)^2}{2\sigma_n^2} \right)dx \nonumber\\ 
&+& 
%\frac{1}{2}\frac{1}{\sqrt{2\pi \sigma_n^2}} 
\left.
\int_{2\delta}^{\infty} \exp\left(-\frac{(x+\delta)^2}{2\sigma_n^2} \right)dx
\right\}. 
\eeqa
On the other hand,
as illustrated in Fig.\ref{fig:16QAMw}, 
the average power of the signal and its distribution are reduced by peak cancellation. 
Since it is clear that the optimum decision threshold for higher order bit is $T^o_H=0$, 
error probability of higher order bit is given as 
\beqa
\hat{P}_{eH} (T^o_H=0)  =  
\frac{1}{2}\frac{1}{\sqrt{2\pi (\sigma_n^2+(\sigma^2_e/\beta))}} \cdot  \nonumber\\
\left\{ \int_{T^o_H=0}^{\infty} \exp\left(-\frac{(x+\mu\delta/\sqrt{\beta})^2}
{2(\sigma_n^2+(\sigma^2_e/\beta))} \right)dx \right.  \nonumber\\
+
\left. \int_{T^o_H=0}^{\infty} \exp\left(-\frac{(x+3\mu\delta/\sqrt{\beta})^2}
{2(\sigma_n^2+(\sigma^2_e/\beta))} \right)dx \right\}. %\nonumber
\label{pH}
\eeqa
Error probability of lower order bit of Gray-coded 16QAM using the optimum decision 
threshold $T^o_L$ in AWGN condition in the case with peak cancellation is 
\beqa
\label{pL}
\hat{P}_{eL} (T^o_L) & = & 
\frac{1}{2}\frac{1}{\sqrt{2\pi (\sigma_n^2+(\sigma^2_e/\beta))}} \cdot \\ \nonumber
& &
\left\{ \int_{-T^o_L}^{T^o_L} \exp\left(-\frac{(x+3\mu\delta/\sqrt{\beta})^2}
{2(\sigma_n^2+(\sigma^2_e/\beta))} \right)dx \right. \\ \nonumber
&+& 
\int_{-\infty}^{-T^o_L} \exp\left(-\frac{(x+\mu\delta/\sqrt{\beta})^2}
{2(\sigma_n^2+(\sigma^2_e/\beta))} \right)dx \\ \nonumber
&+& 
\left.
\int_{T^o_L}^{\infty} \exp\left(-\frac{(x+\mu\delta/\sqrt{\beta})^2}
{2(\sigma_n^2+(\sigma^2_e/\beta))} \right)dx \right\}. 
\nonumber
\eeqa
The optimum decision threshold \textcolor{black}{$T^o_L = -2\mu\delta/\sqrt{\beta}$} to minimize BER 
can be derived by $\frac{\partial P_{eL}(T^o_L)} {\partial T^o_L}=0$. 
Details of the derivation are given in the Appendix. 
Average BER is given by averaging Eq.(\ref{pH}) and Eq.(\ref{pL}).

\subsection{E-SDM MIMO-OFDM}
In this subsection, we consider an Eigen-beam space division multiplexing (E-SDM) OFDM in $M\times N$ MIMO systems, where $M$ and $N$ denotes number of transmit antenna and the number of receive antennas, respectively. 
Hereafter, we assume $M=4$ and $N=2$ as an example. 
PDFs of eigen values with order for random $M\times N$ MIMO channel matrix 
are given as~\cite{12}: 
\beq
f_1(\lambda_1)=\Phi_1(\lambda_1)\exp(-\lambda_1)+\Phi_2(\lambda_1)\exp(-2\lambda_1),
\eeq
\beq
f_2(\lambda_2)=-\Phi_1(\lambda_2)\exp(- 2 \lambda_2),
\eeq
where 
$$
\Phi_1(\lambda)=\lambda^2 \left(\frac{1}{6}\lambda^2 -\lambda +2\right), 
$$
$$
\Phi_2(\lambda)=\lambda^2 \left(\frac{1}{6}\lambda^2 +\lambda +2\right). 
$$
Thus, BER expression of signal transmission over the $i$-th eigen channels 
in $M\times N$ MIMO channel can be derived as
\beq
%P_b^{(1)}=\int^{\infty}_0 f_1(\lambda) \frac{1}{2} Q\left(\sqrt{\frac{\lambda E_b}{N_0} }\right) d\lambda. 
%
P_b^{(i)}=\int^{\infty}_0 f_i (\lambda) \left( \int^{\infty}_0 p(\lambda, x) dx \right) d\lambda, 
\eeq
%Similarly, BER expression for transmission over the second eigen channel can be given as 
%\beq
%P_b^{(2)}=\int^{\infty}_0 f_2(\lambda) \frac{1}{2} Q\left(\sqrt{\frac{\lambda^{(i)} E_b}{N_0} }\right) d\lambda, 
%\eeq
where $M$ and $N$ denote the number of transmit and receive antennas. 
$p(\lambda, x)$ denotes PDF of the signal after peak cancellation in presence of AWGN 
and $\lambda_i$ is the $i$-th eigen vector. 
Similarly single antenna case, for QPSK transmission over $i$-th eigen channel, 
\beq
p(\lambda, x)\!\!=\!\!
%\\ \nonumber
%\frac{1}{\sqrt{ 2\pi (\sigma_n^2 /\lambda + \sigma_e^2 d^2 \lambda) }} 
\frac{1}{\sqrt{ 2\pi (\sigma_n^2 /\lambda + \sigma_e^2 A^2 \lambda) }} 
%%\cdot 
\! \exp \! \left( 
%e^{
-\frac{ (x- \bar{x}_e A\sqrt{\lambda} -A )^2}{ 2( \sigma_e^2 A^2 \lambda + \sigma_n^2 /\lambda 
%-\frac{ (x- \bar{x}_e d\sqrt{\lambda} -d )^2}{ 2( \sigma_e^2 d^2 \lambda + \sigma_n^2 /\lambda
) } 
%}
\right)\!\!. 
\nonumber
\eeq
%where $2d$ denotes distance between two nearest constellation points 
%
%
Average BER over the first and second eigen channel is given as 
\beq
\bar{P}_b = \frac{P_b^{(1)}+P_b^{(2)}} {2}, 
\eeq
where the same power is allocated to the first and the second eigen channels.  
%
%%%Similarly, BER in case of multi-level QAM is derived with the same manner. 
%Although the details on derivation are omitted due to limited pages, 
%theoretical BERs for 16QAM case are shown in next section.
\\

\section{Performance Results and Discussions~\label{section4}}
To clarify the validity of the proposed framework, 
we evaluate the performance of {MIMO-OFDM} system by computer simulation. 
{The system block diagram is shown in the same as in Fig.~\ref{fig:ofdm}.} 
{We assume QPSK, 16QAM, or 64QAM data modulation.} 
The number of FFT points is 512 and the number of subcarriers is 64.
In MIMO cases, Eigen-mode precoding is adopted at each sub-carrier (i.e., E-SDM MIMO), 
where the transmitter and the receiver equip $N$ and $M$ antennas, respectively.
The number of streams is denoted as $K$. Here, we assume $K=N$.  
Channel model is independent attenuated 6-path Rayleigh fading.
%After passing through the channel, AWGN is added to the received signal.  
%
The requirement of ACLR is set to $-50$~dB, 
%pre-defined threshold of ACLR is set to $-50$~dB, 
while those of EVM are set to $-20$, $-25$, and $-30$~dB for QPSK, 16QAM, 
and 64QAM cases, respectively. 
In this paper, we assume that channel estimation is perfectly done at the receiver 
and channel state information is ideally shared with the transmitter. 
%This paper assumes that channel state information is ideally known 
%at both transmitter and receiver end. 

%%
One of the conventional approaches is {repeated C{\&}F}~\cite{19}-\cite{20}. 
The block diagram is shown in Fig.~\ref{CandF}, 
%illustrated in Fig.~\ref{CandF}, 
where $x(t)$ and $x_o(t)$ are the OFDM signals before clipping and after filtering. 
%at the input and 
%the output. 
%$d_c(t)$ is given as the error signal between $x(t)$ and $x_o(t)$};
%$
%d_c(t)=x(t)-x_c(t).
%$
%Here, OoB radiation in the difference signal $d_c(t)$ is 
%suppressed by a low-pass filter (LPF). 
%
%The peak power of the transmit signal is reduced by subtracting 
%the filtered difference signal $d_c(t)$ from the original signal $x(t)$.
%
%in this C\&F, 
To reduce the computational complexity, only the clipped signal (i. e., the peak amplitude exceeds a peak detection threshold) is  band-limited by filtering operation. 
%In this scheme, filtering operation is carried out only when the peak amplitude exceeds a given threshold 
%{in order to reduce its computational complexity}. 
%
%{As illustrated in Fig.~\ref{CandF}, 
%C{\&}F operations are iteratively carried out 
%to suppress the regrowth of signal amplitude caused by filtering, 
%where $N_{it}$ denotes the number of iterations. }
Unlike the proposed peak cancellation, 
this method needs to empirically optimize the filter parameters 
so as to keep EVM and ACLR values below these pre-defined thresholds. 

We also evaluate prior limiter based techniques in~\cite{CandF-3}-\cite{Compand}. 
In~\cite{CandF-3}, the peak detection threshold at $n$-th iteration $A_{th}^{(n)}$ in repeated C\&F 
is adaptively determined as 
\begin{eqnarray}
A_{th}^{(n)} = \sqrt{\frac{N_f}{N_p}}A_{th}^{(n-1)}, 
A_{th}^{(0)} = \frac{1}{N_f}\sum_{s=0}^{N_f-1} |x(s\Delta t)|, 
\end{eqnarray}
where $x(s\Delta t)$ denotes the transmit signal before C\&F operation and 
$\Delta t$ denotes sampling interval. 
$N_f$ and $N_p$ are the number of FFT points and the number of OFDM samples exceeding $A_{th}^{(n-1)}$, respectively. 
In~\cite{Compand}, the following nonlinear function is used to limit peak amplitude:
\begin{eqnarray}
y(s\Delta t) = A_{th} \frac{x(s\Delta t)}{|x(s\Delta t)|} \left( 1+\left( \frac{v}{|x(s\Delta t)|} \right)^{(1/a)} \right)^{-a}, 
\end{eqnarray}
where $x(s\Delta t)$ and $y(s\Delta t)$ are input and output complex time-domain signals of the companding function, respectively. 
$|x(s\Delta t)|$ denotes the absolute value of $x(s\Delta t)$. 
$A_{th}$ is the maximum amplitude of $|y(s\Delta t)|$. $v$ and $a$ are non-linear parameters of the 
nonlinear function, respectively. 

\begin{figure}[t]
	\bcen
	\hspace{10mm}\includegraphics[scale = 0.35]{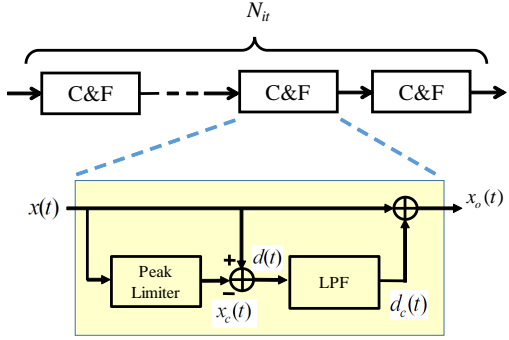}
	\ecen
	\vspace{5mm}
	\caption{Block diagram of repeated C{\&}F, where $N_{it}$ denotes the number of iterations.}
	\label{CandF}
\end{figure}

\begin{figure}[t]
	\begin{center}
		\subfigure[$M$=$N$=1.]{
			\includegraphics[scale=0.45]{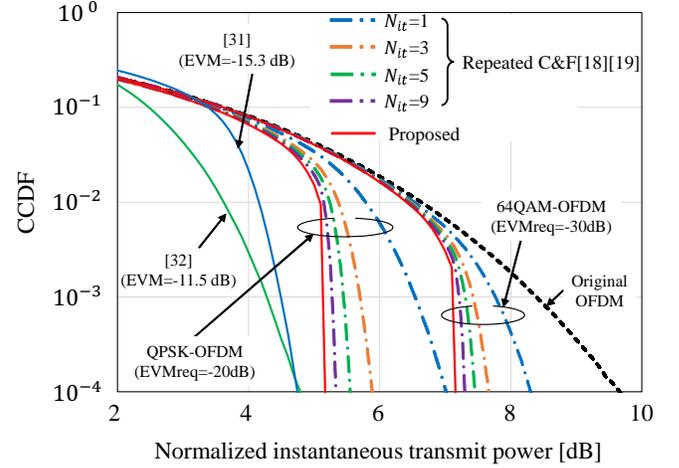} %scale=0.29
		\label{fig:ccdf_siso}}
		\subfigure[$M$=4, $N$=2, $K$=2.]{
			\includegraphics[scale=0.45]{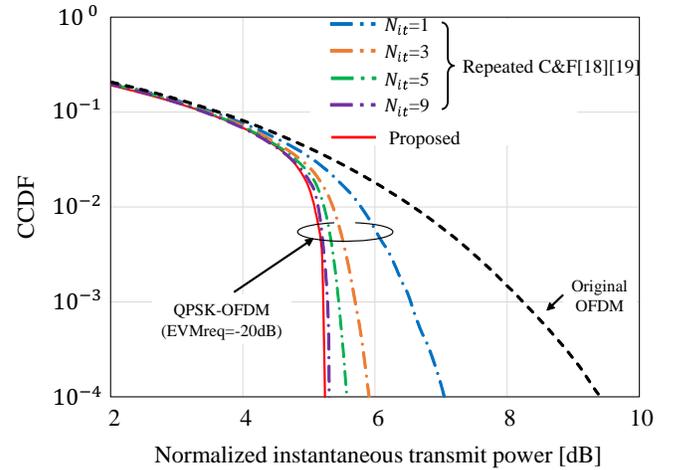} %scale=0.29
		\label{fig:ccdf_mimo}}
		\caption{CCDF of normalized instantaneous transmit power, 
			{where $N_{it}$ denotes the number of iterations in repeated C\&F.}}
		\label{fig:ccdf}
	\end{center}
\end{figure}
%
%\begin{figure}[t]
%	\centering
%	\includegraphics[scale = 0.45]{./fig/Comp_ccdf.eps}
%	\vspace{1mm}
%	\caption{{Comparison of CDDF of proposed method with previoius methods in ~\cite{CandF-3}\cite{Compand}}}
%	\label{fig:comp_ccdf_prev}
%\end{figure}
%
\begin{figure}[t]
	\begin{center}
	\hspace{-5mm}
		\subfigure[QPSK.]{
			\includegraphics[scale=0.42]{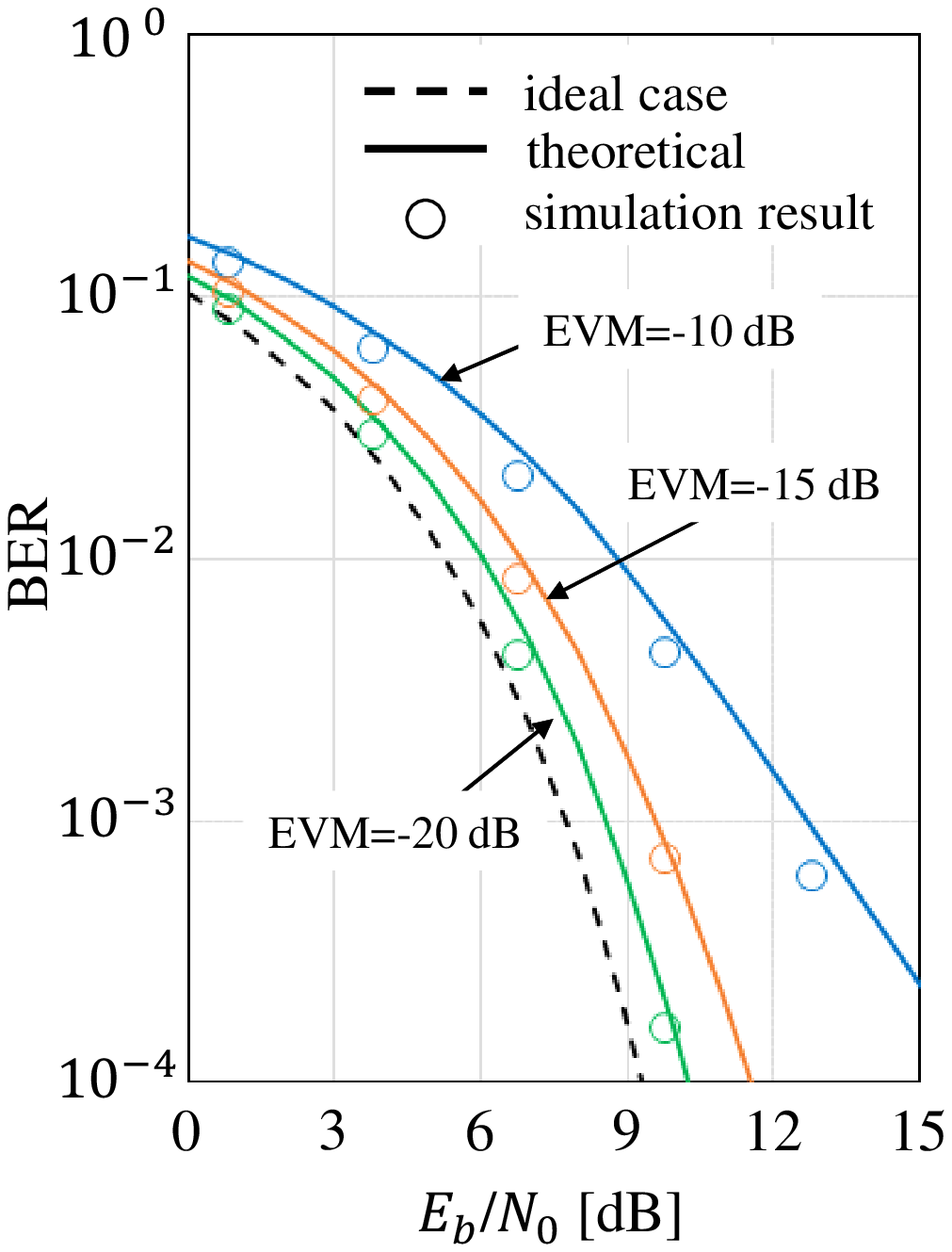} %scale=0.29
		\label{fig:berw_qpsk}}
	\hspace{-8mm}
		\subfigure[16QAM.]{
			\includegraphics[scale=0.42]{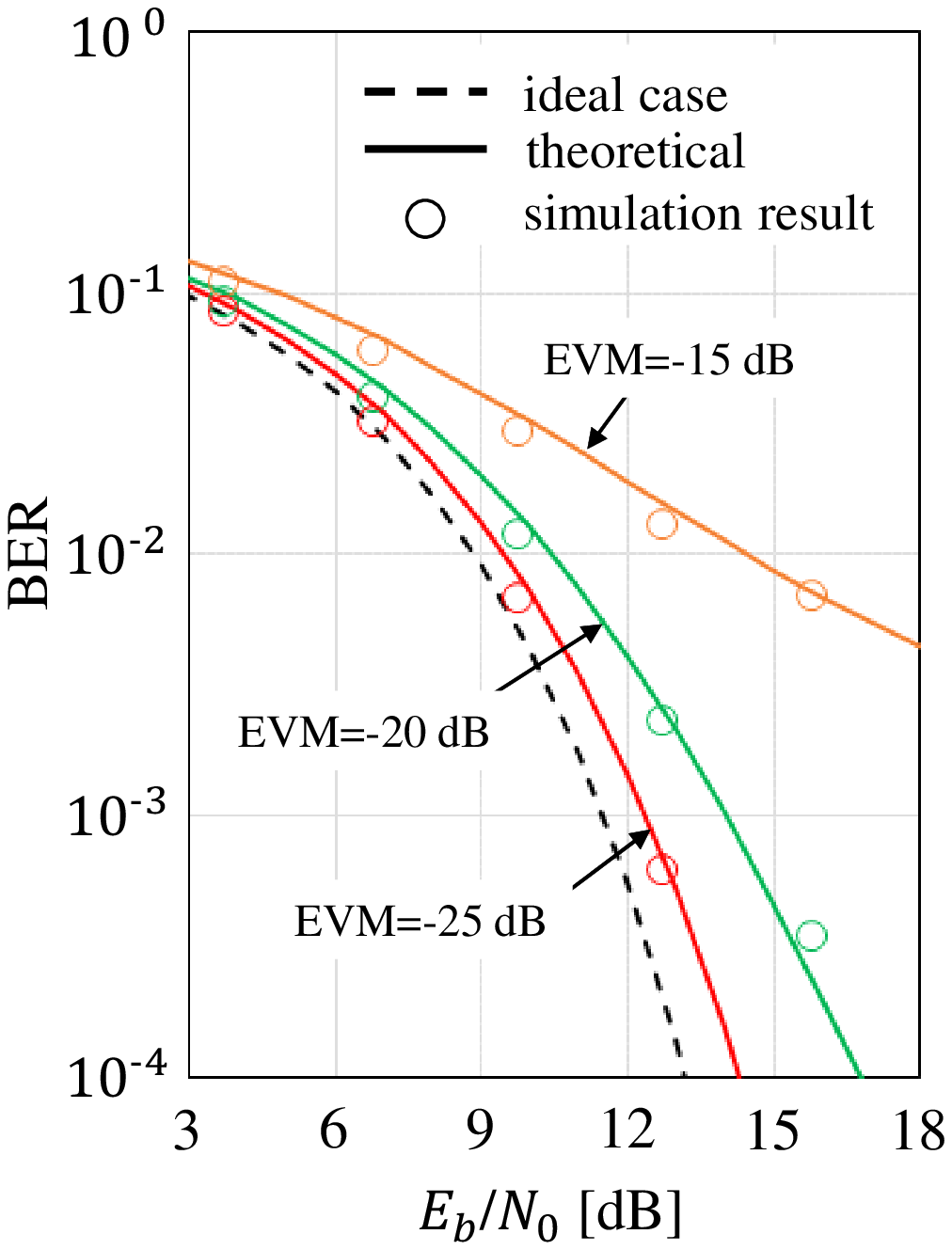} %scale=0.29
		\label{fig:berw_16qam}}
		\caption{Theoretical BER performance and its simulation results of OFDM system using the proposed method in AWGN channel.}
		\label{fig:BER_awgn}
	\hspace{-5mm}
	\end{center}
%\end{figure}
%\begin{figure}[t]
	\begin{center}
	\hspace{-5mm}
		\subfigure[QPSK.]{
			\includegraphics[scale=0.42]{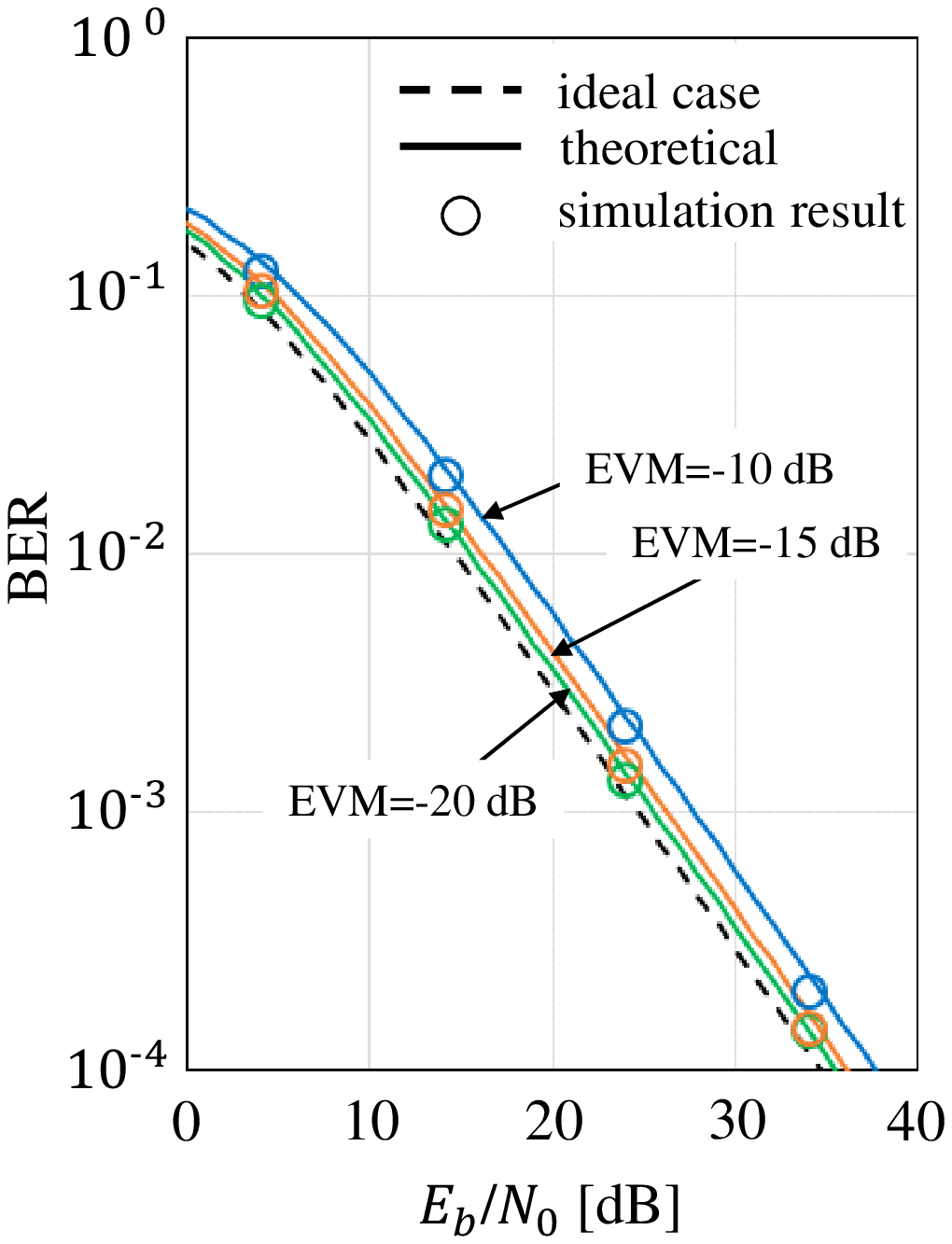} %scale=0.29
		\label{fig:berf_qpsk}}
	\hspace{-8mm}
		\subfigure[16QAM.]{
			\includegraphics[scale=0.42]{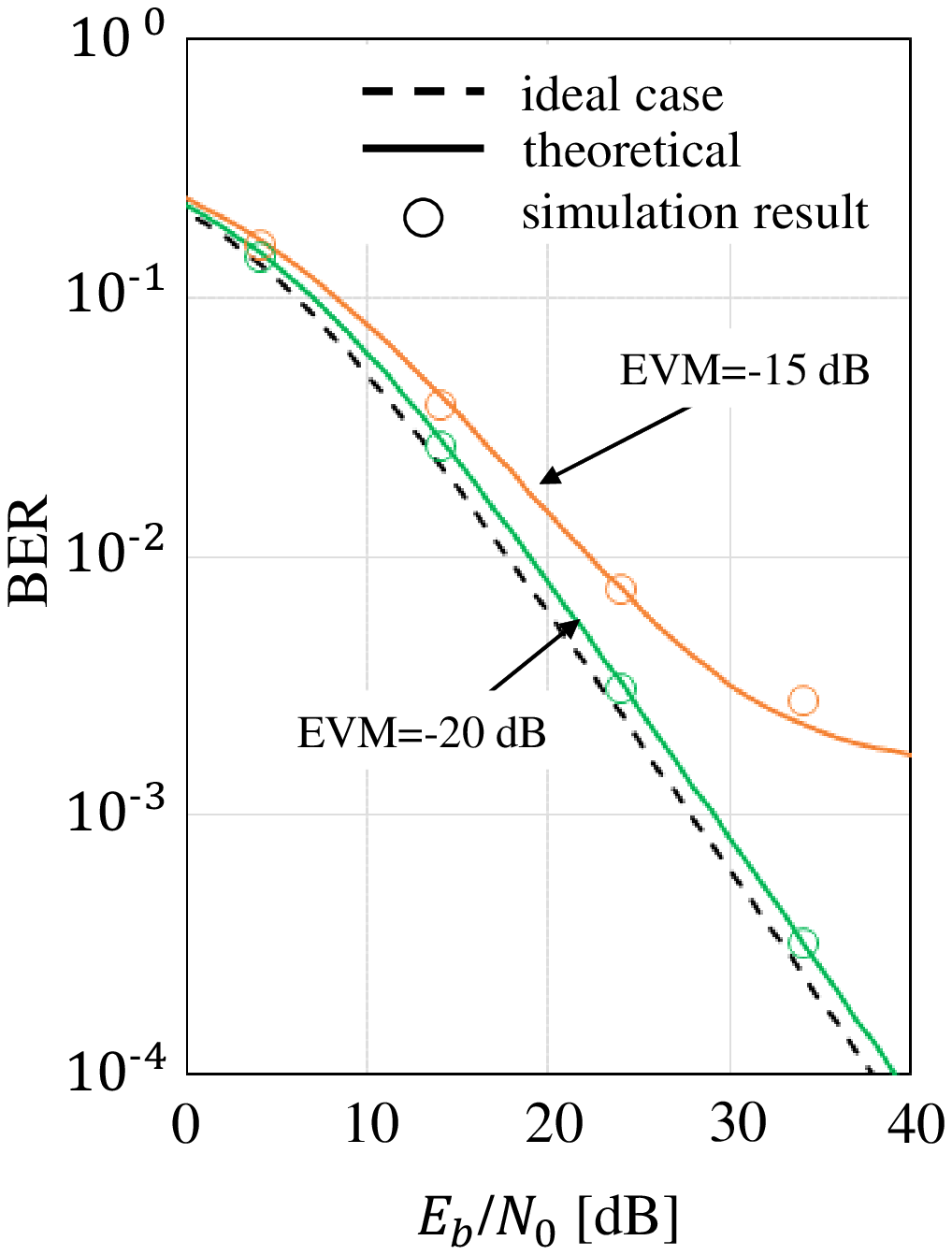} %scale=0.29
		\label{fig:berf_16qam}}
		\caption{{Theoretical BER performance and its simulation results of OFDM system using the proposed method in Rayleigh fading channel.}
	\label{fig:BER_flat}}
	\hspace{-5mm}
	\end{center}
\end{figure}
\begin{figure}[t]
	\begin{center}
	\hspace{-5mm}
		\subfigure[QPSK.]{
			\includegraphics[scale=0.42]{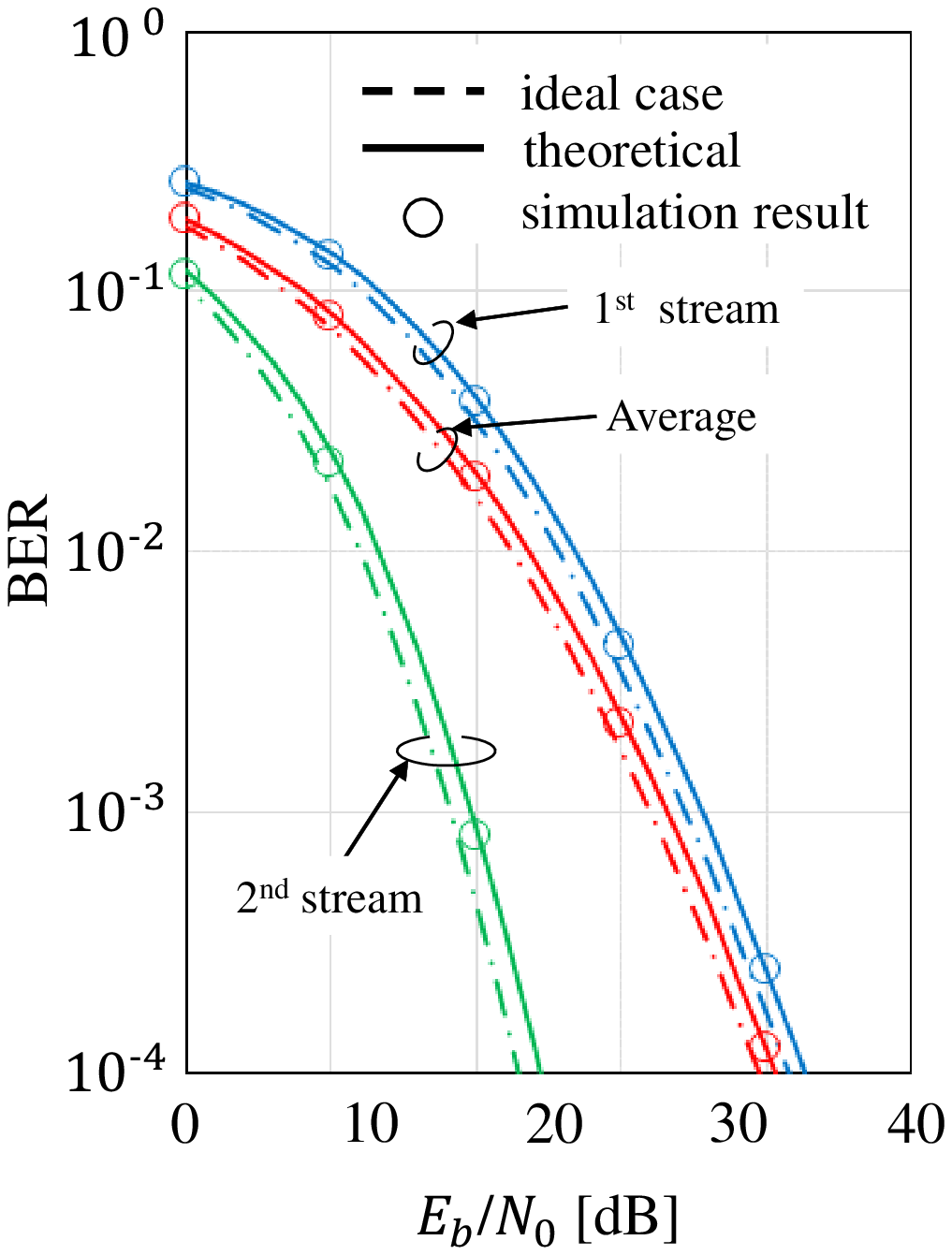} %scale=0.29
		\label{fig:bere_qpsk}}
	\hspace{-8mm}
		\subfigure[16QAM.]{
			\includegraphics[scale=0.42]{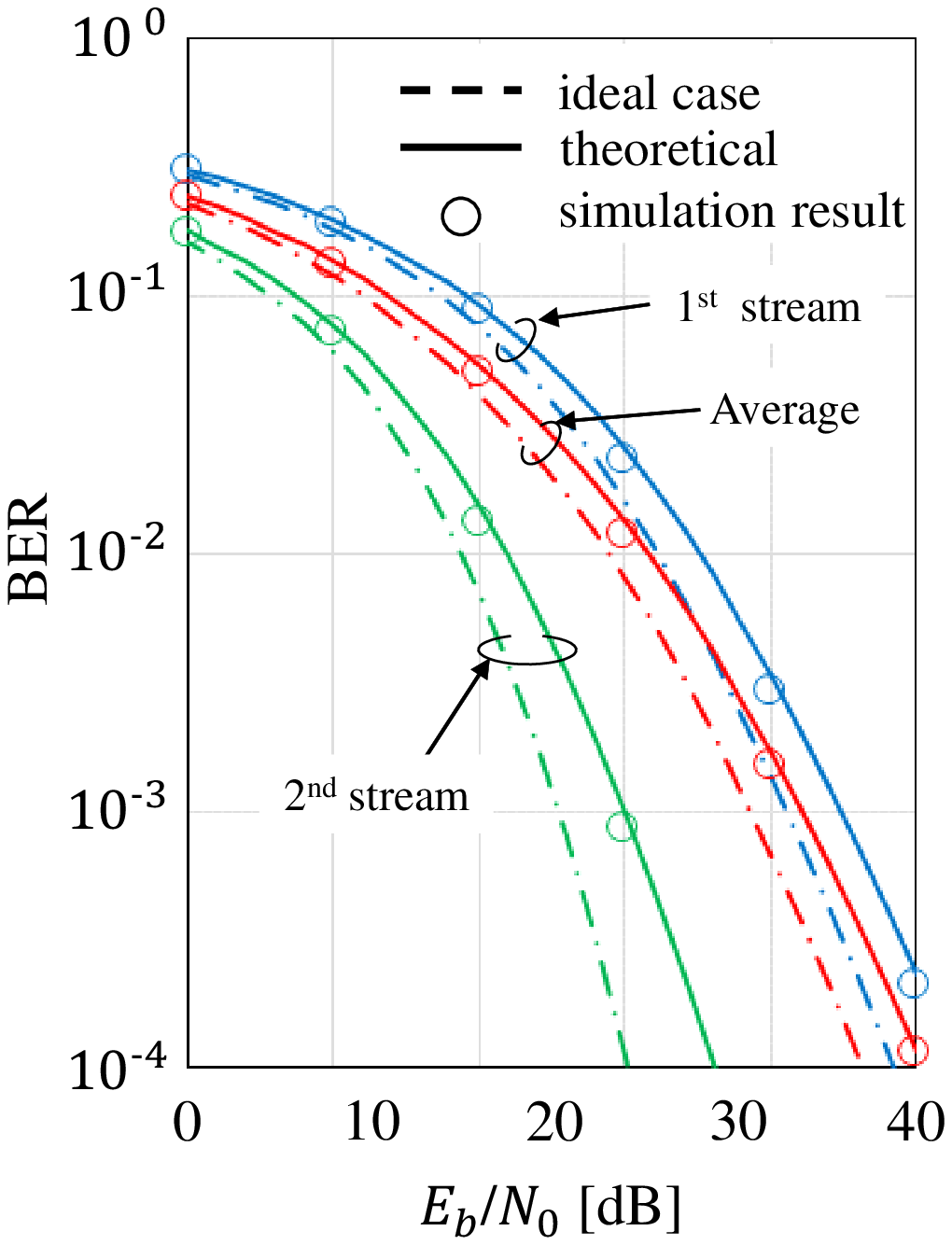} %scale=0.29
		\label{fig:bere_16qam}}
		\caption{{Theoretical BER and its simulation results of E-SDM OFDM system using the proposed method ($N$=4, $M$=2, $K$=2).}}
	\label{fig:BER_ESDM}
	\hspace{-5mm}
	\end{center}
%\end{figure}
%\begin{figure}[t]
\vspace{-6mm}
	\begin{center}
	\hspace{-5mm}
		\subfigure[QPSK.]{
			\includegraphics[scale=0.42]{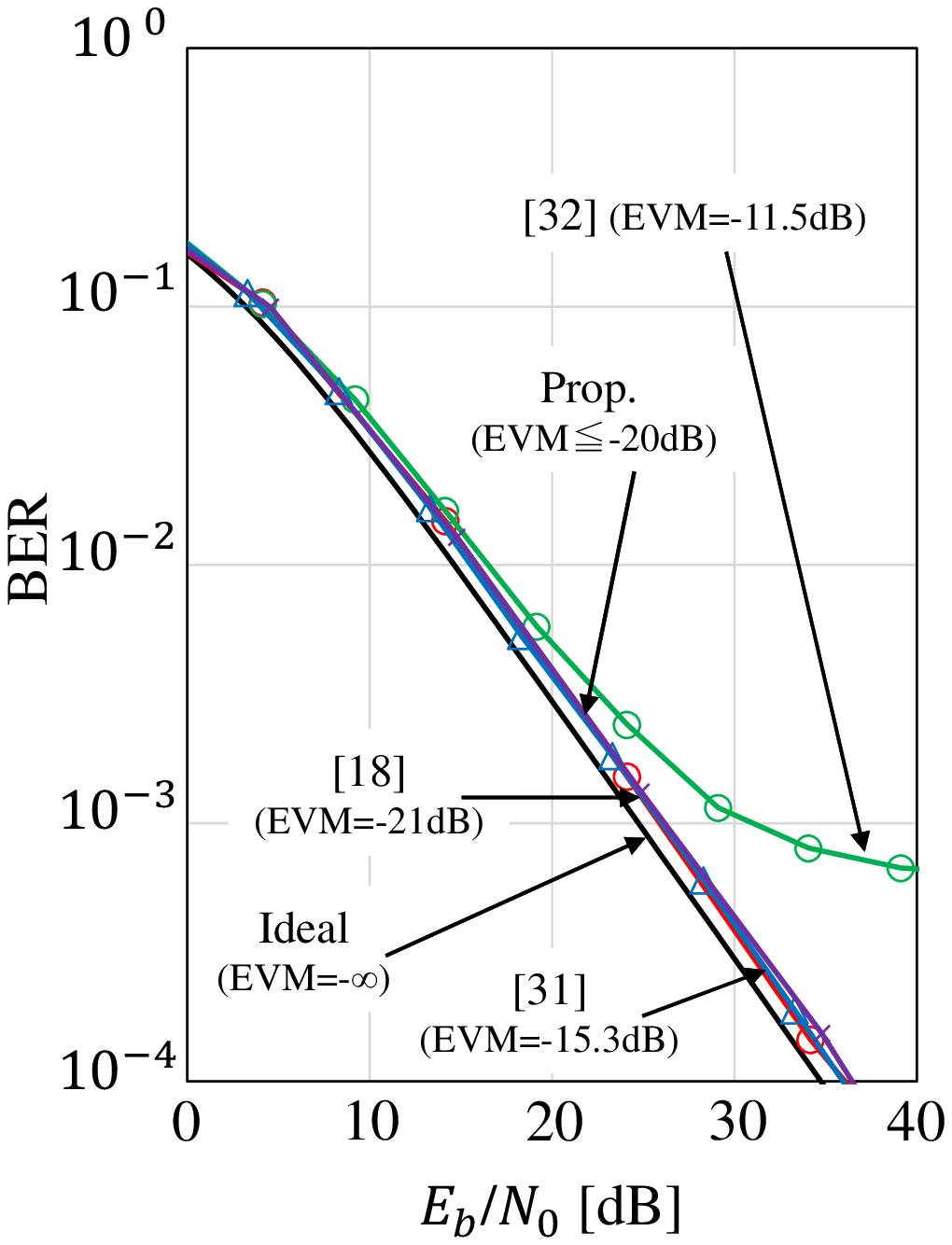} %scale=0.29
			\label{fig:berc_qpsk}}
	\hspace{-8mm}
		\subfigure[16QAM.]{
			\includegraphics[scale=0.42]{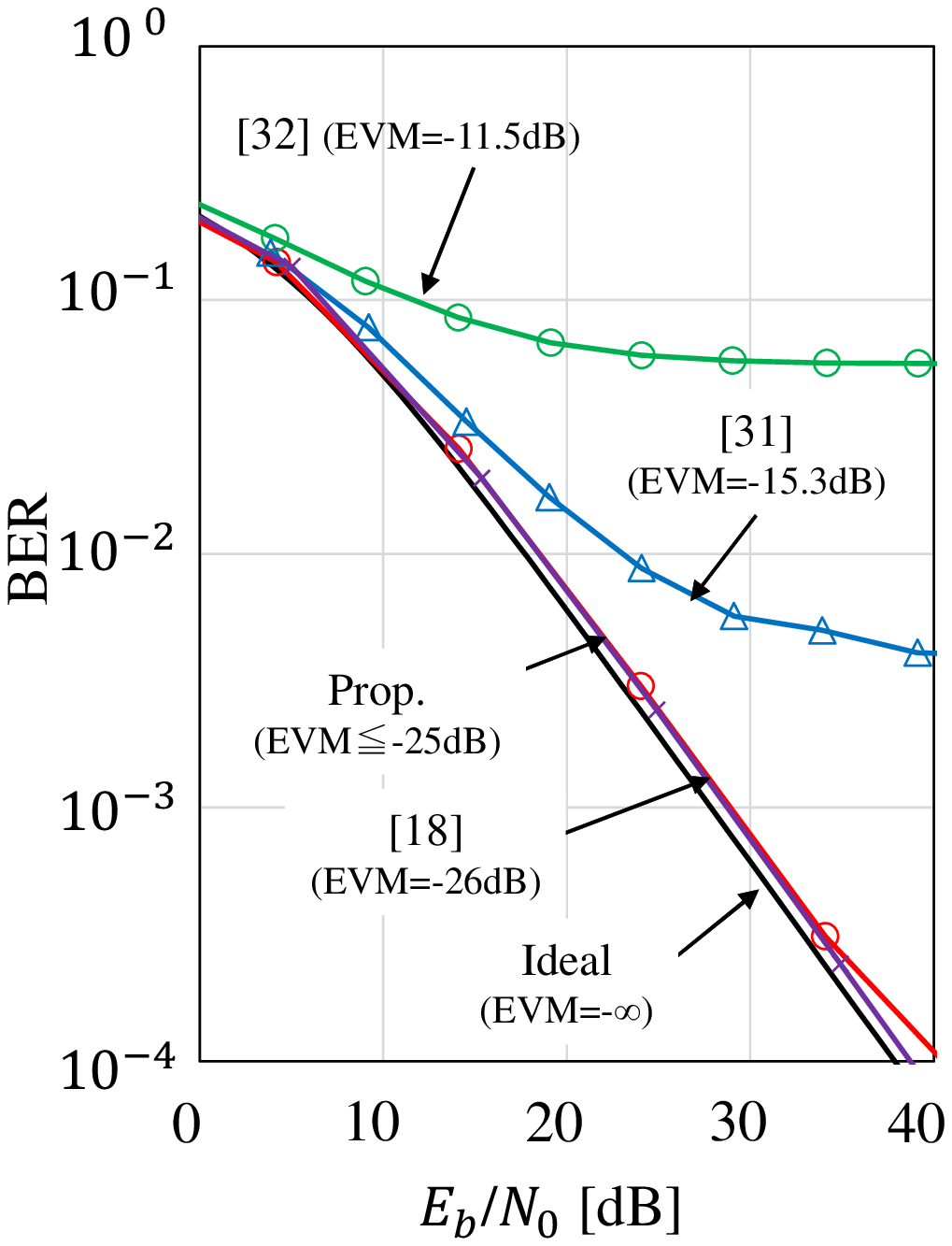} %scale=0.29
			\label{fig:berc_16qam}}
		\caption{\textcolor{black}{BER comparison of the proposed method with conventional approaches in Rayleigh fading channel, where QPSK and 16QAM are used as subcarrier modulation, respectively.}}
		\label{fig:BER_comp}
	\hspace{-5mm}
	\end{center}
\end{figure}

\subsection{PAPR}
Figures~\ref{fig:ccdf_siso} and (b) illustrate the 
statistical distribution of the normalized instantaneous signal power 
of transmitting signals in single-antenna OFDM and $4\times2$ MIMO-OFDM, respectively. 
Here, we use CCDF to evaluate the statistical distribution  
and the  instantaneous is normalized by average power of the signal. 
For comparison,
CCDFs of cases of repeated C\&F in~\cite{19}\cite{20} and other limiter based approaches in~\cite{CandF-3}\cite{Compand} are also shown. 
In repeated C\&F~\cite{19}\cite{20}, we note here that both ACLR and EVM satisfy the required values. 
%are kept below pre-defined values. 
%
In~\cite{Compand}, $v$=7, $a$ = 0.05 and $A_{th}$ = 1 are used. 
In~\cite{CandF-3}\cite{Compand}, roll-off filter with roll-off factor = 0 is used. 
Figure~\ref{fig:ccdf_siso} shows that the instantaneous power of transmit signal 
with the proposed method is limited to around 5.12~dB and 7.25~dB 
at CCDF=$10^{-4}$ for QPSK and 64QAM data modulation schemes. 
The optimum threshold value is $P_{th}$=5.12~dB and 7.25~dB for QPSK and 64QAM data modulation schemes, respectively. 
Note that the achieved PAPR by the proposed method is close to the optimum threshold value. In other words, the threshold value corresponds to the achievable PAPR. 
As for comparison with other approaches, although 
methods in~\cite{CandF-3}\cite{Compand} show lower PAPR
than the proposed 
method, it suffers from high in-band distortion which results in BER degradation as discussed in 
Fig.~\ref{fig:berc_16qam}. Note that the proposed method is able to reduce the PAPR to a lower value compared to~\cite{CandF-3}\cite{Compand} if the required EVM value is set to a higher value, because the proposed method works to automatically minimize the peak power under given EVM and ACLR requirements. 
%Fig.~\ref{fig:berc_16qam}. 
It can be also seen that instantaneous power in the case of repeated C\&F is approaching 
the proposed method by increasing the number of iterations in the repeated C\&F; 
almost the same CCDF in comparison with the proposed method is achieved with $N_{it}=9$. 
Similar findings are observed in Fig.~\ref{fig:ccdf_mimo} for E-SDM case. 

\subsection{BER}
Figures~\ref{fig:berw_qpsk} and (b) illustrate the BER performance 
of single-antenna OFDM with the peak cancellation in AWGN channel, 
In this figure, peak cancellation is conducted under ACLR and EVM restrictions. 
Here, QPSK and 16QAM are used as subcarrier modulation. 
The BER curve of OFDM without any degradation due to peak cancellation
is labeled as "ideal case".
It is demonstrated in Fig.~\ref{fig:BER_awgn}(a) that the proposed method achieves very close BER performance to ideal case, where EVM satisfies the pre-defined requirements $-20$ dB for QPSK and $-25$ dB for 16QAM, respectively. 
These figures also indicate that the theoretical BER curves show good agreements with its simulation results. 

Figure~\ref{fig:BER_flat} and ~\ref{fig:BER_ESDM} illustrate the BER performance 
of single-antenna OFDM and $4\times2$ MIMO using eigen-mode in Rayleigh fading channel, respectively. 
Here QPSK and 16QAM are used as subcarrier modulation. EVM and ACLR are restricted below the pre-defined values, respectively.
These figures also indicate that the theoretical BER curves show good agreements with their simulation results. 
It can also be seen from Fig.~\ref{fig:BER_ESDM} that 
BER performance of the first and second streams in E-SDM scenarios show good agreements with their simulation results. 
We note here that EVM and ACLR meet the pre-defined values.

\textcolor{black}{BER of the proposed method is compared with those of the conventional approaches in~\cite{19}, \cite{CandF-3} and~\cite{Compand} in Fig.~\ref{fig:BER_comp}, 
where QPSK and 16QAM are used for subcarrier modulation in Figs.~\ref{fig:berc_qpsk} and Fig.~\ref{fig:berc_16qam}, respectively.
The Purple line shows the BER of the repeated C\&F~\cite{19} using the same threshold as the proposed peak cancellation.} 
The blue line and the green line show the BER of the repeated C\&F in~\cite{CandF-3} and that of the companding 
technique in~\cite{Compand}, respectively. 
Here, parameters in the repeated C\&F and companding are same in Fig.~\ref{fig:ccdf_siso}. 
The red line shows the BER of the proposed method. 
\textcolor{black}{
In Fig.~\ref{fig:BER_comp}, for QPSK case, it can be seen that the proposed method and conventional methods in~\cite{19}\cite{CandF-3} 
show almost the same BER.  
On the other hand, for 16QAM case, BER of the conventional methods are significantly degraded, 
while the proposed method shows good BER performance comparable to the ideal case.
This is because the proposed method is able to keep EVM below the predefined threshold automatically unlike the conventional
methods.}
\\

\subsection{Complexity}
This subsection evaluates a computational complexity of required PAPR reduction procedures between the two methods above. 
In this comparison, 
the number of complex multiplications is used as the complexity metric. 
%, while the number of complex additions is not taken into consideration for brevity. 
%
%The computational complexity of the proposed method and repeated C{\&}F method using time domain filtering are expressed as
When PC signal is added $N_{pc}$ times per OFDM symbol, 
the complexity of the proposed method is $\left\langle N_w \times N_{pc} \right\rangle$, 
where 
$N_{w}$ denotes the number of complex multiplications per PC addition.
% and 
%$N_{pc}$ stands for the number of PC signal additions per OFDM symbol.
%
On the other hand, 
the complexity of the C{\&}F with time domain LPF is given as  
$\left\langle \sum_{j=1}^{N_{it}} (N_{tap} \times N^{(j)}_{th}) \right\rangle$. 
%\begin{equation}
%\hspace*{-3mm}
%C_{o}=
%\left\{
%\begin{array}{ll}
%
%\left\langle N_w \times N_{pc} \right\rangle & \mbox{Proposed method} \\
%
%\left\langle \sum_{j=1}^{N_{it}} (N_{tap} \times N^{(j)}_{th}) \right\rangle & \mbox{C{\&}F (Time domain LPF)},
%\end{array}
%\right.
%\nonumber
%\end{equation}%
Here, $\left<x\right>$ denotes the averaging of $x$. 
$N_{it}$ and $N_{tap}$ denote the number of iterations and the number of taps in time domain filter, respectively.
{$N^{(j)}_{th}$ is the number of samples exceeding a given threshold 
per OFDM symbol. 
$j$ denotes the iteration index. }

\begin{figure}[t]
\centering
\hspace*{5mm}\includegraphics[scale = 0.44]{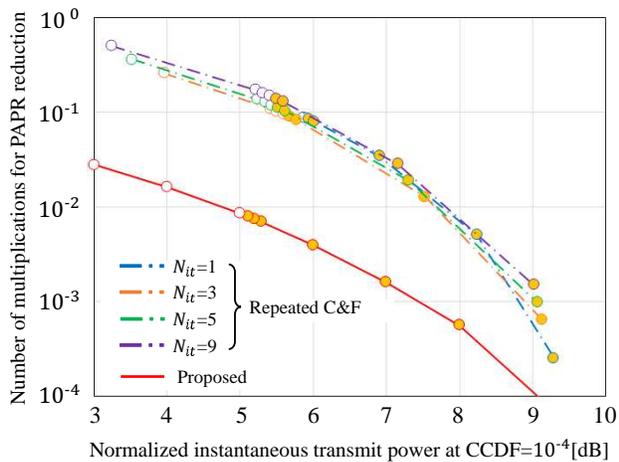}
\vspace{1mm}
\caption{{Number of multiplications per OFDM symbol for PAPR reduction 
($M=4$, $N=2$, $K=2$), where QPSK is adopted as subcarrier modulation.}}
\label{fig:comp}
\end{figure}

The required complexity for the peak cancellation in SDM-OFDM is evaluated in comparison with the repeated C{\&}F 
based on the above defined metric in Fig.~\ref{fig:comp} where $M=4$ and $N=2$ are assumed. 
In the figure, 
we assume QPSK is adopted as subcarrier modulation. 
Here, PAPR is defined as the normalized instantaneous power observed at CCDF=$10^{-4}$. 
{The red line shows the result of the proposed method, 
while the results of repeated C\&F method are depicted by other color lines.} 
The color-fill circular shape markers represent 
that both EVM and ACLR requirements are fulfilled, 
while the white-fill shows that either EVM or ACLR (or both) exceed the pre-defined value. 
In repeated C\&F cases, 
normalized instantaneous power is approaching the peak detection threshold 
$S_{th}=10\log_{10}\frac{A^2_{th}}{S_t}$ by repeating clipping and filtering operations. 
Note that in the proposed method, ACLR and EVM can be kept below the required values automatically, unlike the repeated C\&F method. 
%Note that the proposed method keeps ACLR and EVM below the pre-defined values automatically, unlike the repeated C\&F method. 
%
This figure proves that the required complexity for the proposed peak cancellation 
is lower in comparison with {the repeated C{\&}F}.

\begin{figure}[t]
\centering
\hspace*{5mm}\includegraphics[scale = 0.28]{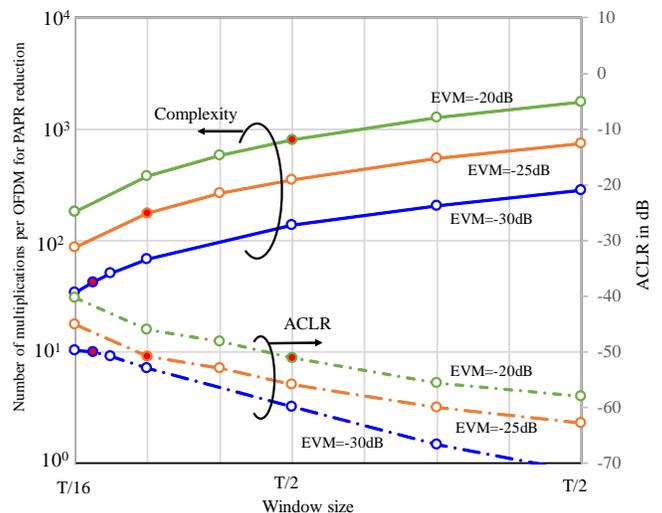}
\vspace{-6mm}
\caption{{Relationship between complexity and ACLR in terms of required EVM, 
where peak detection threshold values are set to optimum values to meet EVM requirements.}}
\label{fig:comp_aclr}
\end{figure}

\subsection{Windowing}

{Figure~\ref{fig:comp_aclr} illustrates the relationship between the required complexity 
and achievable ACLR as a function of window size truncating the PC signal 
in terms of required EVM values which are set to $-20$, $-25$ or $-30$~dB, 
where red drawn circle markers denote selected window size for each EVM value 
in case of ACLR$=-50$~dB.
The threshold is selected to meet the required EVM as discussed in Sect.~\ref{section3c} 
and correspond to achievable PAPR and BER. 
The PC signal is truncated by the window function $w(t)$ 
as shown by Eqs.~(\ref{pcsignal})-(\ref{window}), 
Thus, window-size $T_2$ affects both ACLR and the complexity. 
It can be seen that from this figure as window-size increases, ACLR decreases 
at the expense of increasing the required complexity.
{The results} suggest that the required complexity is minimized by 
optimizing the window-size under a given EVM (i.e., BER) and threshold value (i.e., PAPR). }
\\

%\subsection{Comparison with other PAPR reduction methods}
\section{Conclusion~\label{section5}}
Performance analysis of a dynamic peak-cancellation scheme for E-SDM OFDM system 
has been presented, where EVM and ACLR are automatically restricted below a pre-defined level. 
%
%In this method, the maximum amplitude is replicated with a PC signal and the peak amplitude of transmit signal is canceled out with the replicated one. 
%
Using the proposed approach, degradations due to ACLR and EVM are 
effectively mitigated while keeping the OoB radiation below its target value. 
Furthermore, practical design of peak-cancellation signal is discussed with target OoB radiation and in-band distortion through optimizing the windowing size of the PC signal.
In addition, we have also theoretically analyzed peak cancellation capability and achieved BER, respectively. 
Numerical results prove that our peak cancellation method is effective in improving both the BER and PAPR under EVM and ACLR restrictions. 
It can be also seen that theoretical BER show good agreements with simulation results. 
\clearpage
\textcolor{black}{
\appendix
The derivation of $T_L^o$ in Eq.~(\ref{pL}) is as follows. 
Without loss of generality, we assume $\beta$=1. 
Since Gray-encoded 16QAM is used, the error probability of lower- order bits is given as 
\begin{eqnarray}
\hat{P}_{eL} (T_L) & = & 
\frac{1}{2}\frac{1}{\sqrt{2\pi (\sigma_n^2+(\sigma^2_e))}} \cdot \nonumber \\
& &
\left\{ \int_{-T_L}^{T_L} \exp\left(-\frac{(x+3\mu\delta)^2}
{2(\sigma_n^2+(\sigma^2_e))} \right)dx \right. \nonumber \\
&+& 
\int_{-\infty}^{-T_L} \exp\left(-\frac{(x+\mu\delta)^2}
{2(\sigma_n^2+\sigma^2_e)} \right)dx \nonumber \\
&+& 
\left.
\int_{T_L}^{\infty} \exp\left(-\frac{(x+\mu\delta)^2}
{2(\sigma_n^2+\sigma^2_e)} \right)dx \right\} \nonumber \\
&=& 1 + {\rm erf} \left( \frac{T_L+3\mu\delta}{\sqrt{2(\sigma_n^2+\sigma_e^2)}}\right) \nonumber \\
&-& {\rm erf} \left( \frac{T_L+\mu\delta}{\sqrt{2(\sigma_n^2+\sigma_e^2)}}\right),
\end{eqnarray}
where $T_L$ is the threshold level to decide I and Q phase of 16QAM constellation points. 
Using the relationship
\begin{eqnarray}
\int \frac{1}{\sqrt{2\pi\sigma_n^2}}\exp\left( -\frac{x^2}{2\sigma_n^2}\right)dx = 
\frac{1}{2}\left( 1 + {\rm erf}\left( \frac{x}{\sqrt{2\sigma_n^2}}\right)\right), 
\end{eqnarray}
the partial differentiation of $\hat{P}_{eL} (T_L)$ with respect to $T_L$ is given as 
\begin{eqnarray}
\frac{\partial\hat{P}_{eL} (T_L)}{\partial T_L} &=& \frac{1}{\sqrt{2(\sigma_n^2+\sigma_e^2)}} \cdot \nonumber \\
& &\left( \exp\left( \frac{(T_L+3\mu\delta)^2}{\sqrt{2(\sigma_n^2+\sigma_e^2)}}\right) \right. \nonumber \\
&-&\left. \exp\left(\frac{(T_L+\mu\delta)^2}{\sqrt{2(\sigma_n^2+\sigma_e^2)}}\right)\right). 
\end{eqnarray}
By solving $\frac{\partial\hat{P}_{eL} (T_L^o)}{\partial T_L^o} = 0$, 
\begin{eqnarray}
\exp\left(\frac{(T_L^o+3\mu\delta)^2}{\sqrt{2(\sigma_n^2+\sigma_e^2)}}\right) &=& \exp\left(\frac{(T_L^o+\mu\delta)^2}{\sqrt{2(\sigma_n^2+\sigma_e^2)}}\right) \nonumber \\
(T_L^o+3\mu\delta)^2 &=& (T_L^o+\mu\delta)^2 \nonumber \\
T_L^o &=& -2\mu\delta, 
\end{eqnarray}
where $-2\mu\delta$ is an intersection point of two Gaussian distributions located on the left side of the horizontal axis of 16QAM constellation as shown in Fig.~\ref{fig:16QAMw}.}

\section*{Acknowledgment}
This research was supported in part by 
%the Support Center for Advanced Telecommunications Technology Research, Foundation, 
JSPS KAKENHI (JP17K06427, JP17J04710), 
and 
Kyushu University Short-term International Research Exchange Program 
by University Research Administration Office. 
\\

\end{document}